%% file: main_IOP.tex
\documentclass[12pt]{iopart}
\usepackage{iopams}
\usepackage[nolist]{acronym}
\usepackage{pdfpages}
\usepackage{multirow}
\usepackage{tabto}
\usepackage{floatrow}
\usepackage{graphicx}
\usepackage{subfig}
\usepackage{wrapfig}
\usepackage{xcolor}
\usepackage{multicol}
\usepackage[colorlinks]{hyperref}
\usepackage[capitalize]{cleveref}
\usepackage{siunitx}
\usepackage{esint}
\usepackage{upgreek}

\usepackage[normalem]{ulem}
\usepackage{comment}
\usepackage{multicol}

\newcommand{\gls}{\ac}
\newcommand{\Gls}{\Ac}
\newcommand{\glspl}{\acfp}

\newcommand{\muup}{\upmu}
\newcommand{\scaletable}[1]{\scalebox{0.75}{#1}} 

\newcommand{\lopdeq}{L_{\mathrm{OPD}}}
\newcommand{\xoff}{x_\textrm{off}}
\newcommand{\xcor}{x_\textrm{cor}}
\newcommand{\zoff}{z_\textrm{off}}
\newcommand{\zcor}{z_\textrm{cor}}

\begin{document}

\newacro{HG}{Hermite-Gaussian}
\newacro{IFO}{interferometer}
\newacro{GW}{gravitational wave}
\newacro{LISA}{Laser Interferometer Space Antenna}
\newacro{TMI}{test mass interferometer}
\newacro{LA IFO}{long arm interferometer}
\newacro{IFOs}{interferometers}
\newacro{TTL}{tilt-to-length} 
\newacro{SC}{spacecraft}
\newacro{TM}{test mass}
\newacro{MB}{measurement beam}
\newacro{LO}{local oscillator}
\newacro{IS}{imaging system}
\newacro{LPS}{longitudinal pathlength signal}
\newacro{PD}{photodiode}
\newacro{QPD}{quadrant photodiode}
\newacro{CoR}{center of rotation}
\newacro{LPSA}{averaged phase}
\newacro{LPST}{total signal}
\newacro{LPSC}{complex sum}
\newacro{DWS}{differential wavefront sensing signal}
\newacro{DPS}{differential power sensing signal}
\newacro{GA}{general astigmatic}
\newacro{CoM}{centre of mass}
\newacro{GRS}{gravitational reference system}
\newacro{OB}{optical bench}
\newacro{MEM}{mode expansion method}
\newacro{OPD}{optical pathlength difference}
\newacro{ISI}{inter-satellite interferometer}
\newacro{GRS}{gravitational reference system}

\title{Impact of phase signal formulations on tilt-to-length coupling noise in the LISA test mass interferometer}

\author{Paul Edwards$^{1*}$, Megha Dave$^2$$^,$$^3$, Alexander Weaver$^1$,\\ Mengyuan Zhao$^4$, Paul Fulda$^1$, Guido Mueller$^1$$^,$$^2$$^,$$^3$,\\ and Gudrun Wanner$^3$$^,$$^2$}

\address{$^1$ University of Florida Physics Department, University of Florida Department of Physics, 2001 Museum
Road, Gainesville, Florida 32611, USA}
\address{$^2$ Max Planck Institute for Gravitational Physics (Albert-Einstein-Institute),
Callinstr. 38, D-30167 Hannover, Germany}
\address{$^3$ Institute for Gravitational Physics of the Leibniz Universit{\"a}t Hannover,
Callinstr. 38, D-30167 Hannover, Germany}
\address{$^4$ Key Laboratory
of Electronics and Information Technology for Space System, National Space Science
Center, Chinese Academy of Sciences, No.1 Nanertiao, Zhongguancun, Haidian district,
Beijing 100190, China}

\ead{paul.edwards@ufl.edu}

\vspace{10pt}
\begin{indented}
\item[]February 2025
\end{indented}


\begin{abstract}
\Gls{TTL} coupling in the \gls{LISA} can generate spurious displacement noise that potentially affects the measurement of gravitational wave signals. 
In each test mass interferometer, the coupling of misalignments with spacecraft angular jitter produces noise in the longitudinal pathlength signal. 
This signal comes from the phase readout combinations of an interfering measurement beam and a local reference beam at a quadrant photodetector, but various formulations exist for its calculation. 
Selection of this pathlength signal formulation affects the \gls{TTL} coupling noise. 
We therefore simulated two pathlength signal candidates, the complex sum and the averaged phase, and evaluated their performance in the \gls{LISA} test mass interferometer under the assumption of static alignment imperfections.
All simulations were performed using three different methods with cross-checked results.
We find with all three methods that the averaged phase is the choice with equal or less \gls{TTL} coupling noise in four out of the five test cases.
Finally, we show that the non-linear \gls{TTL} contributions in the test mass interferometer to be negligible in all our test-cases, no matter which formulation of pathlength signal is chosen.
\end{abstract}

\vspace{2pc}

%
\maketitle
%
%


\input{S1_Introduction}

\input{S2_Simplification}

\input{S3_Methods}

\input{S4_Results}

\input{S5_Conclusions}

\section{Acknowledgments}
We thank Josep Sanjuan for providing his valuable insight and the LISA Consortium's TTL Expert Group as well as Gerhard Heinzel for inspiring discussions. \\
This work was supported by NASA grant 80NSSC19K0324 for contributions of the team in Florida, US. The Albert Einstein Institute acknowledges support of the German Space Agency, DLR and by the Federal Ministry for Economic Affairs and Climate Action based on a resolution of the German Bundestag (FKZ 50OQ0501, FKZ 50OQ1601 and FKZ 50OQ1801). Additionally, Gudrun Wanner acknowledges funding by Deutsche Forschungsgemeinschaft (DFG) via its Cluster of Excellence QuantumFrontiers (EXC 2123, Project ID 390837967).
Furthermore, Gudrun Wanner and Megha Dave acknowledge support by the DFG Cluster of Excellence PhoenixD (EXC 2122, Project ID 390833453) and thank both clusters for the opportunities and excellent scientific exchange via the Topical Group ``Optical Simulations'' and the Research Area ``S'' on Simulations. \\
This work has also been supported by the Max Planck Society and the Chinese Academy of Sciences (CAS) in the framework of the LEGACY cooperation on low-frequency gravitational wave astronomy (M.IF.A.QOP18098, CAS’s Strategic Pioneer Program on Space Science XDA1502110201). The National Key R\&D Program of China (2020YFC2200100) provided the funding that supported the contributions made by Mengyuan Zhao in this work.

\clearpage

\appendix
\input{A1_square_qpd_check}


\section*{References}
\bibliographystyle{iopart-num}
\bibliography{references}

\end{document}

%% file: S1_Introduction.tex
\section{Introduction}

The Laser Interferometer Space Antenna (\gls{LISA}) is a gravitational wave detector that was recently adopted by ESA with an anticipated launch in the mid-2030s \cite{LISA_adoption}. 
The primary objective of \gls{LISA} is to observe the gravitational wave spectrum in the millihertz band, a range that is currently inaccessible to ground-based detectors \cite{LISA_L3}.
This band is richly populated by compact binary systems in which the characteristic strain is $\mathcal{O}(10^{-21})$.
\gls{LISA} will consist of a triangular constellation of three identical spacecraft, each featuring two movable optical sub-assemblies that point along the line-of-sight to the other two remote spacecraft. 
Each sub-assembly houses a telescope and an optical bench with several interferometers for science and reference measurements. 
Each spacecraft is equipped with two \glspl{GRS}, comprising the hardware and avionics necessary to hold, release, and maintain the respective test masses in free-fall. A specialized electrode housing within a \gls{GRS}, which features capacitive position sensing capabilities, encloses a single free-falling Au-Pt alloy test mass{, monitoring}{. This monitors} any displacement of the test mass from its center.

To enable length measurements between three pairs of test masses, six one-way laser links will be established between the spacecraft.
The test mass-to-test mass differential length fluctuations generated by gravitational waves - and by certain noise sources - will be measured at each optical bench via the heterodyne beats of a \gls{MB} with a reference beam, termed the \gls{LO}, which enters from the adjacent optical bench of the spacecraft. 
For technical reasons, the test mass-to-test mass link is split into three individual measurements. 
The first is the \gls{ISI}, which measures the distance between a link's optical benches forming the endpoints of the inter-spacecraft distance. 
The other two measurements are at the \glspl{TMI} of linked spacecraft, which measures displacement of each test mass with respect to the local OB.
The \gls{TMI}, which is the focus of this work, is sketched in \cref{fig:tm_ifo_aligned} below.

\begin{figure}[hbp!]
\centering
\includegraphics[width=.8\textwidth,
    trim=2.5in .5in 2in .5in,
    clip]{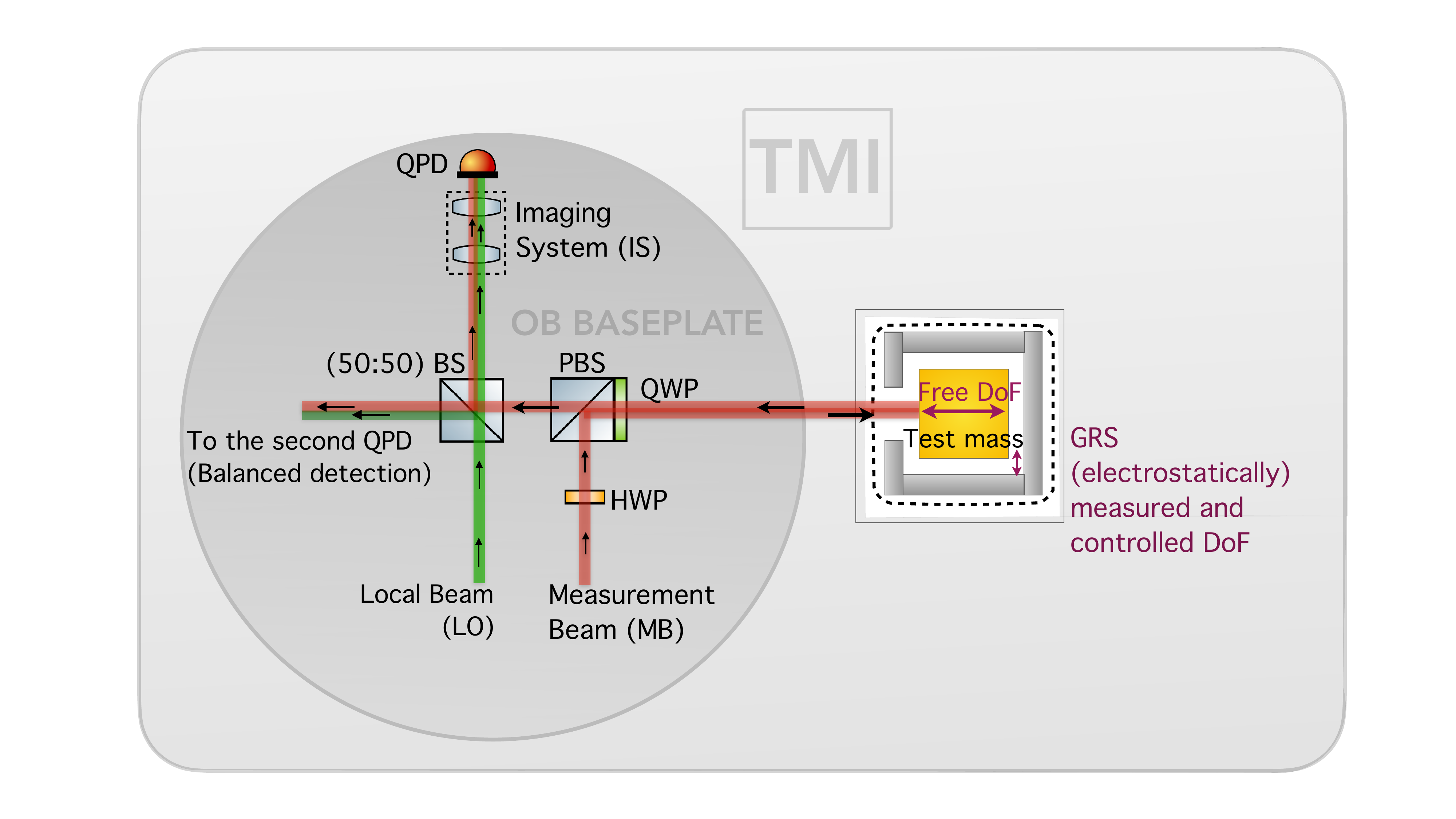}
\caption{\label{fig:tm_ifo_aligned}A condensed sketch of the test mass interferometer. 
Upon reflection from the test mass, the measurement beam is combined with the local oscillator at a beamsplitter before passing through an imaging system and being detected at a quadrant photodetector.}
\end{figure}

By utilizing post-processing techniques and combining interferometer readouts, \gls{LISA} will achieve its required $\textrm{pm}/\sqrt{\textrm{Hz}}$ displacement sensitivity over the mission bandwidth. 
The feasibility of this pm-scale sensitivity has been partly proven by the LISA Pathfinder (LPF) mission \cite{LPF}.
However, several noise sources inherent to LISA's interferometers remain to be investigated. 
For one, the interferometric detectors are vulnerable to spurious displacement fluctuations, e.g. due to spacecraft angular jitters. 
This leakage of angular motion into the pathlength signal is known as \gls{TTL} coupling, and was already shown to be a significant noise source in LPF \cite{TTL_in_LPF}. 

There are numerous publications on \gls{TTL} coupling summarized, for example, in \cite{Wanner2024}. 
However, only one of these has considerable overlap with the content presented here.
Of all the publications on \gls{TTL} coupling, a recent paper (\cite{ttl_suppress_imaging_system}) by members of the Taiji community, has the closest connection to ours with a study of the \gls{TTL} noise in the \gls{TMI}. 
However, that paper discusses the optimization of the design of an imaging system and its placement for the suppression of first and second order geometric \gls{TTL} within the \gls{TMI} of the Taiji mission, as well as in-orbit calibration.
Likewise, results in \cite{Hartig_2022,Hartig_2023} consider cross-coupling from lateral jitter of the \gls{TM}.
Meanwhile, our study is focused on investigating various static misalignments coupled with spacecraft-to-test mass angular jitter at the interferometric \gls{QPD} level, effectively maintaining a black box view of the imaging optics.
In particular, we demonstrate in this paper that the way in which the longitudinal pathlength signal (LPS) itself is defined can serve to preemptively mitigate \gls{TTL} noise caused by certain misalignments. 
This arises from the fact that the \gls{QPD} utilized in the \gls{TMI} generates four distinct pathlength signals, each corresponding to a detector quadrant.
Although several definitions have been proposed to combine these four signals into a unified pathlength readout \cite{WANNER20124831, wanner_readouts}, no definitive selection has been made for the \gls{TMI} yet.
So, a key question in optimizing readout remains: which pathlength definition is least affected by \gls{TTL} contributions?

To investigate pathlength formulations and identify patterns in signals generated by particular noise sources,  we have systematically probed a parameter space of misalignments which would generate \gls{TTL} noise in the \gls{TMI} readout.
These misalignments can be inherited by the \gls{MB} and \gls{LO} at the \gls{QPD}, so we simulated interference of the misaligned beams and computed their beat signals.
We then used these signals to compare the readout fidelity of two specific longitudinal pathlength formulations - \gls{LPSA} and \gls{LPSC} - for their use at the \gls{TMI}.
The remainder of the paper reports this simulation and is structured as follows:
\begin{itemize}
    \item \cref{Section2} provides an overview of TTL coupling in the TMI and a full description of our simulation parameter space.
    \item \cref{Section3} contains a summary of the three simulation methods and a derivation of the interferometric signal formulations.
    \item \cref{Section4} is focused on results of the two signal formulations and the effectiveness of eliminating the \gls{TTL} coupling by subtracting the linear fit model from the two signal formulations.
    \item Finally, \cref{Section5} concludes with implications of the results and outlines future work involving TTL noise simulation for the \gls{ISI}.
\end{itemize}

%

%% file: S2_Simplification.tex
\section{Simulation at the test mass interferometer}\label{Section2}

In this section, we explore potential sources of pathlength readout noise in the \gls{TMI}. 
We introduce a simplified simulation environment that distills an otherwise complex setup, including the imaging system, to its fundamental components and focuses on testable parameters related to \gls{TTL} noise.
From there, we outline the parameter space of our simulations.

\subsection{The imaging system and tilt-to-length coupling}
\label{sub:imaging_ttl}

Each \gls{TMI} comprises a free-falling test mass in an electrode housing, several optics, and a \gls{QPD}. Additionally, a Gaussian \gls{MB} propagates through components on the optical bench, to the test mass, then finally to the \gls{QPD} where it combines with a Gaussian \gls{LO}.
However, in its path from the test mass reflective surface, as in \cref{fig:tm_ifo_aligned}, the \gls{MB} acquires pathlength fluctuations with respect to the reference \gls{LO} due to relative spacecraft-to-test mass angular jitter. 

Imaging optics mitigate these fluctuations by reorienting each beam's final \gls{CoR} - ideally, to the center of the \gls{QPD}. 
Ultimately, the true success of the imaging system lies in its cancellation of the phase offset present between the \gls{LO} and the misaligned \gls{MB} on their combination at the \gls{QPD}. 
This phase rectification capability can be described most simply with a single, combined lens, as in \cref{fig:IS_wavefront} below.
%
\begin{figure}[!htbp]
    \centering
    {
    \includegraphics[width=.95\textwidth,
    trim=.8in .5in 1.3in .in,
    clip]{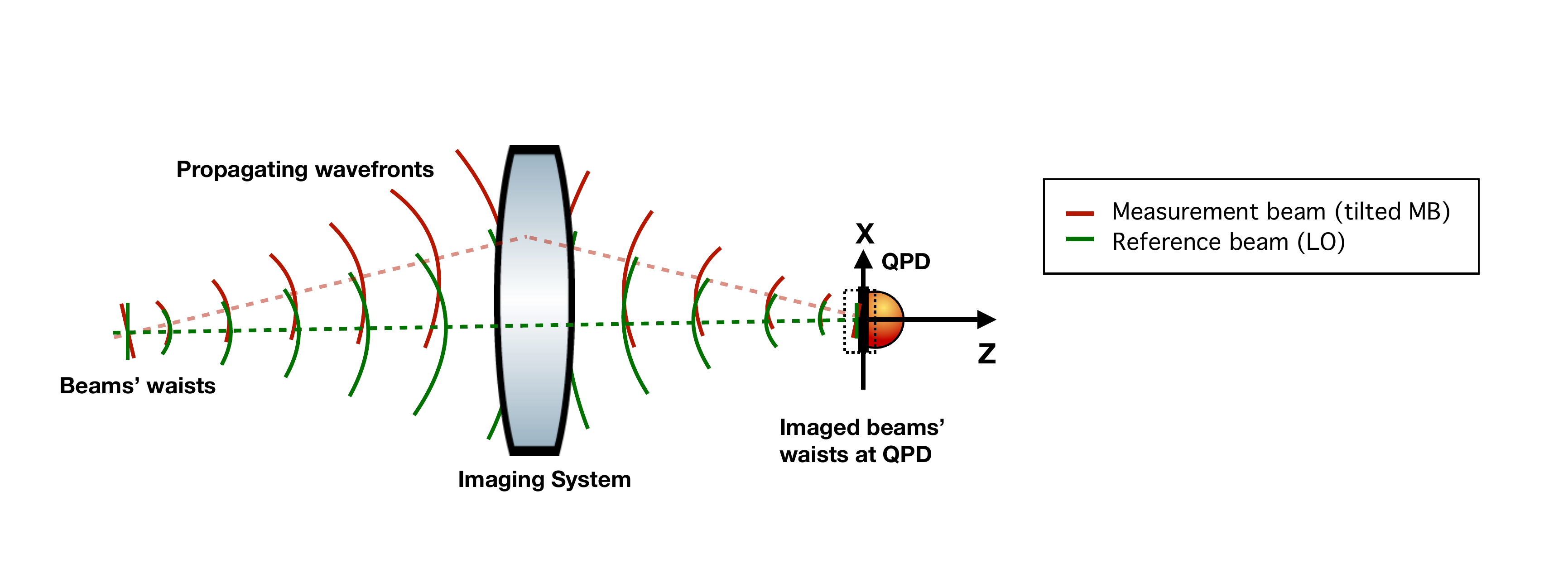}}
        \caption{This simplified sketch of the imaging system depicts: a single lens; a measurement beam (the dashed optical axis in red); a local oscillator (the dashed optical axis in green); and a photodiode, which has a detector surface centered at the origin of the axes.
    The propagating wavefronts are imaged to ensure their phase offset, which otherwise introduces a spurious pathlength signal, is cancelled on combination at the detector. 
    }
    \label{fig:IS_wavefront}
\end{figure}
\subsection{Simplification of the simulation environment}
\label{sub:nom_setup}

\begin{figure}[!hbp]
    \centering \includegraphics[width=.95\textwidth,
    trim=0.3in 5.5in 0.5in 3.5in,
    clip]{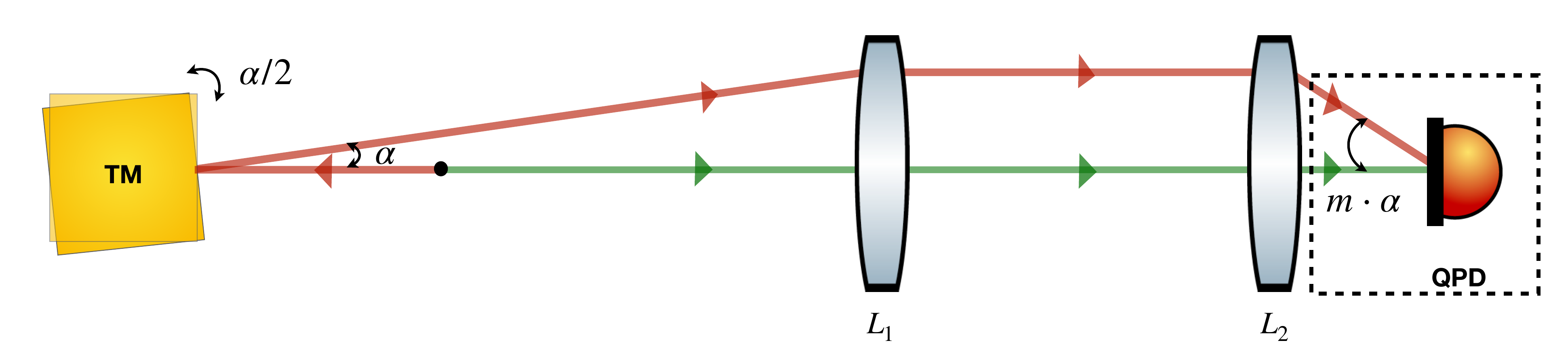} 
    \caption{
    A sketch of the condensed test mass interferometer, where spacecraft-to-test mass relative jitter yields a beam tilt $\alpha$ in the \gls{MB} (red) relative to the \gls{LO} (green). After passing the imaging system, shown here with two lenses, the \gls{MB} is redirected onto the \gls{QPD} at an angle, but it is recentered at the \gls{QPD}.
    }
    \label{fig:e}
\end{figure}

Assuming perfect imaging without abberation, the imaging system can most easily be incorporated into simulation by considering its geometric effect on beam angles and sizes. 
Firstly, the \gls{IS} alters the beam angle from $\alpha$ before the \gls{IS}, to $(\alpha')$ after the \gls{IS}, and both angles relate by the magnification factor $m$:
\begin{equation}
\label{eq:1}
\alpha' = m \cdot \alpha  \;,    
\end{equation}
and we define both angles relative to the QPD normal (cf.\cref{fig:e}).
%
Meanwhile, the incident beam waist ($w_0$) is scaled to the emergent beam waist ($w_0'$) by the magnification as
\begin{equation}
\label{eq:2}
w_0' =  \frac{w_0}{m} \;.
\end{equation}
Here, we assume that the waists are located in the pupil planes of the \gls{IS}.
Ideally, the \gls{IS} then images a center of rotation which is located along the beam axis prior to the imaging system, onto the center of the \gls{QPD}. The geometric effect observable on the \gls{QPD} is given by a simple scaling of the incident beams and a rotation of the \gls{MB} about the \gls{QPD} center. Under these circumstances, the imaging optics suppress all geometric TTL coupling.
Nevertheless, non-geometric effects can cause residual TTL noise to arise, e.g., from mismatches of the interfering wavefront curvatures.
This can be mitigated by matching the beam properties, such that the two beam waists share an identical size and location on the \gls{QPD} surface. In such a case, the non-geometric TTL also vanishes, as shown in \mbox{\cite{Schuster15_vanishing_ttl}} and illustrated in \mbox{\cref{fig:IS_wavefront}}. 

In our simulation, the goal was to consider the dominant \gls{TTL} noise sources upstream along the \gls{TMI}, before beam combination, and condense them into a simple model.
The \gls{TMI} is a complex system of optics and other components which are not necessary to implement in a full \gls{TTL} simulation.
Instead, pathlength signals are more simply derived from a combined parameter space of perturbations that we consider to be inherited by the \gls{MB} and \gls{LO} in their final relative geometries at the \gls{QPD}.
Therefore, a more simple simulation environment is one which includes only the dashed box portion of \cref{fig:e} - that is, the misaligned combining beams and the \gls{QPD}. 
This is the testbed used in our simulation.

By this simplification, we must propagate all misalignments through the beam geometries such that angular jitter is translated into beam misalignments or offsets with respect to the nominal optical axes.
However, if we consider the dominant sources of misalignment to be the test mass and \gls{IS}, as in \cref{fig:e}, we need only to incorporate their geometric effects into the parameter space to simulate practical interferometric signals.

\subsection{Simulation setup}
\label{Sec:SimSetup}
Due to the described simplification, we assume a condensed setup to describe the key features of the \gls{TMI}. 
The setup features are nominally:

\begin{itemize}

    \item The \gls{MB} and \gls{LO} have the same beam properties, i.e., equal waist sizes and Rayleigh ranges ($z_\textrm R$).
    \item The waists of the \gls{MB} and \gls{LO} are scaled down with the magnification of the imaging system ($m=2.5$).
    \item For the non-rotated case of the \gls{MB}, the \gls{MB} and \gls{LO} have identical beam axes, which impinge at normal incidence onto the center of the \gls{QPD}. 
    The waist locations and the \gls{MB}'s center of rotation nominally lie at the center of the \gls{QPD} representing the case of ideal imaging.
    \item The simulated jitter of the \gls{MB} is twice the expected jitter of the test mass relative to the spacecraft (see \cref{fig:e}). Additionally, this doubled angle is further magnified by the imaging system, such that the total beam jitter on the \gls{QPD} is five times that of the spacecraft-to-test mass angular jitter.  
    \item The waist size of each beam is set to a conservative value ($\sim 400\,\muup$m) such that it fits sufficiently on the $1\,$mm-width \gls{QPD} active area.
\end{itemize}

Based on the assumptions outlined above, we consider an ideal scenario characterized by the absence of both geometric and non-geometric \gls{TTL} coupling (see \cref{fig:e,fig:IS_wavefront}). 
In the context of this study, we analyze the impact of coupling on two signal formulations. 
We specifically focus on deviations from this ideal case, which we generically term misalignments.

\subsection{Beam, photodiode, and misalignment parameters}
\label{sub:param_space}

The previous section described the nominal settings. 
We now describe the imperfections and variations to these nominal settings that we consider as outlined by \cref{params_table}. 
The nominal values here represent the zero \gls{TTL} case of ideal alignment of the \gls{MB} and \gls{LO} along the optical axis. These are listed in the second column.
The third column shows the ranges for which we tested the \gls{TTL} coupling caused by variations of the listed parameters, and the fourth column shows the subsection in which the corresponding test is described.
\begin{table}[htbp!]
    \caption{\label{params_table}Nominal and test values for parameters used in  simulation.}
    \scaletable{
    \begin{tabular}{@{}llll}
    \br
          Parameter  & 
          Nominal Value & 
          Test Values & 
          Subsections\\
    \mr
          Wavelength $\lambda$ (both beams) &
          1064\,nm &
          -&
          -\\
  
    \mr
          Circular Photodiode Diameter &
          1\,mm &
          - &
          \ref{sn:pd_dim}\\
    \mr
          Square Photodiode Width &
          \{$\sqrt{2}$/2, 1\}\,mm & 
          -   & 
          \ref{sn:pd_dim}\\
    \mr
        Photodiode Gap &
        20\,$\muup$m &
        - & 
        \ref{sn:pd_dim}\\
    \mr
          Beam Tilts ($\alpha$) &
           0\,$\muup$rad & 
           [-900, 900]\,$\muup$rad  &
         \ref{sn:beam_tilt}\\
    \mr
                Waist radius ($w_0$) &
          402\,$\muup$m  & 
          \{362,382,402,422,442\}\,$\muup$m &
          \ref{sn:beam_size}, \ref{sub:w0_diff}\\
    \mr
            Beam Lateral Offset ($x_{\textrm{off}}$) &
            $0\,$mm & $\pm \{0, 0.1, 0.2\}\,$mm &
            \ref{MB_Intro_Lat}, \ref {sub:x_off}\\
    \mr
            Beam Lateral CoR ($x_{\textrm{cor}}$) &
            0\,mm &
            $\pm \{0, 0.5, 1\}$\,mm &
            \ref{COR_Intro_Lat}, \ref{sub:x_cor}\\
    \mr
          Beam Waist Longitudinal Offset ($z_{\textrm{off}}$) &
            0\,mm &
             $ \pm\{0, 23.8, 47.6\}$\,mm &
             \ref{MB_Intro_Long}, \ref{sub:z_off}\\
    \mr
           Beam Longitudinal CoR ($z_{\textrm{cor}}$) &
            0\,mm &
            $ \pm \{0, 2.5, 5\}$\,mm &
            \ref{COR_Long_Rot}, \ref{sub:z_cor}\\
    \br
    \end{tabular}
    }
\end{table}
Note that small changes in both the longitudinal and lateral positions of the beams and \gls{CoR} could significantly increase the noise \cite{Hartig_2023}. For example, in reality, the \gls{CoR} of the spacecraft (or the \gls{MB}) is some arbitrary point that yields a longitudinal and lateral offset at the entrance pupil of the lens system. 
Similarly, some errors are expected in the position of the beams.
The following subsections illustrate the simulation parameters and further describe their selection criteria under this consideration.

\subsubsection{Photodiode dimensions} \label{sn:pd_dim}

\begin{figure}[htbp!]
\centering
    \subfloat[]{\includegraphics[width=.5\textwidth,
    trim=1.5in 2.5in .1in 1.5in,
    clip]{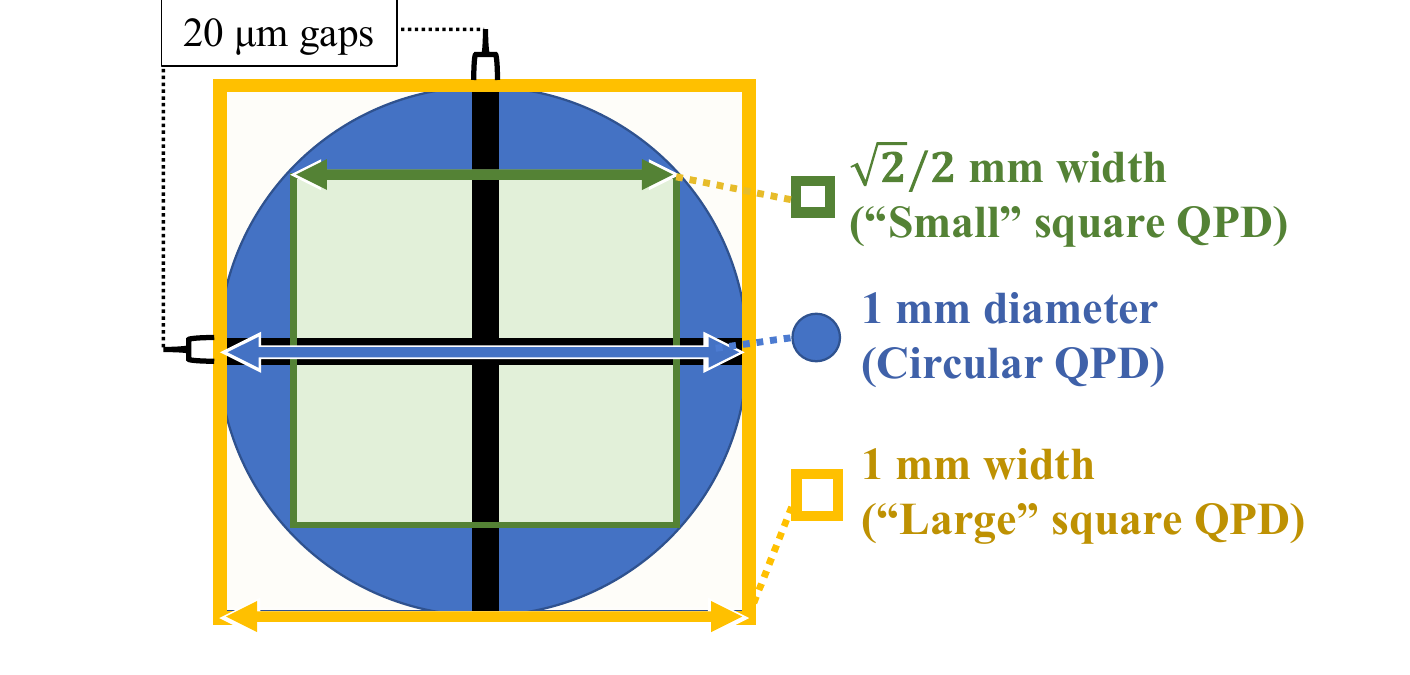}\label{subfig:beam_qpd}}
   \subfloat[]{\includegraphics[width=.35\textwidth]{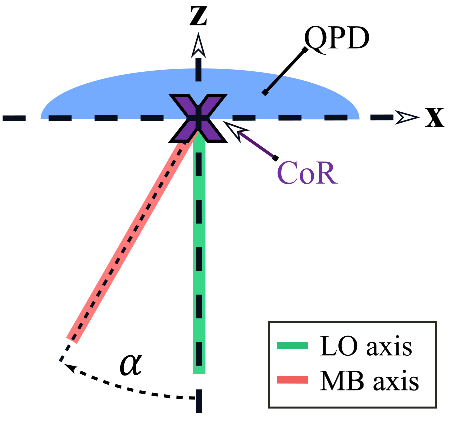}\label{subfig:sketch_default}}
    \caption{(a) shows overlays of the 1 mm-diameter circular (blue), $\sqrt{2}/2\,$mm ``small'' square (green), and 1 mm ``large'' square (yellow) QPDs used in simulation.
    In all test cases, inactive gaps between quadrants were 20 $\muup$m (not to scale here). 
    (b) shows the simplest misalignment case - \gls{MB} (red) rotation by angle $\alpha$ about the QPD center (purple ``X''), with an on-axis \gls{LO} (green) and no other parameter variations.}
\end{figure}

In \gls{LISA}, we expect circular quadrant photodiodes to be used and assume for the dimensions of these a diameter of $1\,$ mm and a gap width of $20\,\muup$m \cite{LISA_L3,pd_cervantes_2011, qpd2024}.
However, circular geometry can be implemented in only one of the three simulation methods used in this paper.
For this reason, we also tested square \gls{QPD}s, which can be used with all three methods.
To test the effect of the square shape, we implemented two different sizes for these square \gls{QPD}s: ``small''  $\sqrt{2}/2\,$mm-width squares, which are inscribed within the circular \gls{QPD}, and ``large'' 1 mm-width square which are circumscribed around the circular \gls{QPD}s.
These settings are sketched in Fig.~\ref{subfig:beam_qpd}.  

\subsubsection{Beam tilt}\label{sn:beam_tilt}
The ``default'' geometry of beam tilt ($\alpha$) with no other misalignment is shown in \cref{subfig:sketch_default}.
Note that the \gls{LO} (shown in green) is optimally aligned along the interferometric $z$-axis, and the only variation to the \gls{MB} (shown in red) is a rotation around the $y$-axis and about the \gls{QPD}'s center.
For this setup, we tested how much the beam could be tilted before the interferometric contrast vanishes. Additionally, we tested the linearity of the DWS signal over the same range.

\begin{figure}[!htbp]
\label{fig:dws_and_contrast}
\centering
    \sidesubfloat[]{\includegraphics[width = .45\textwidth]{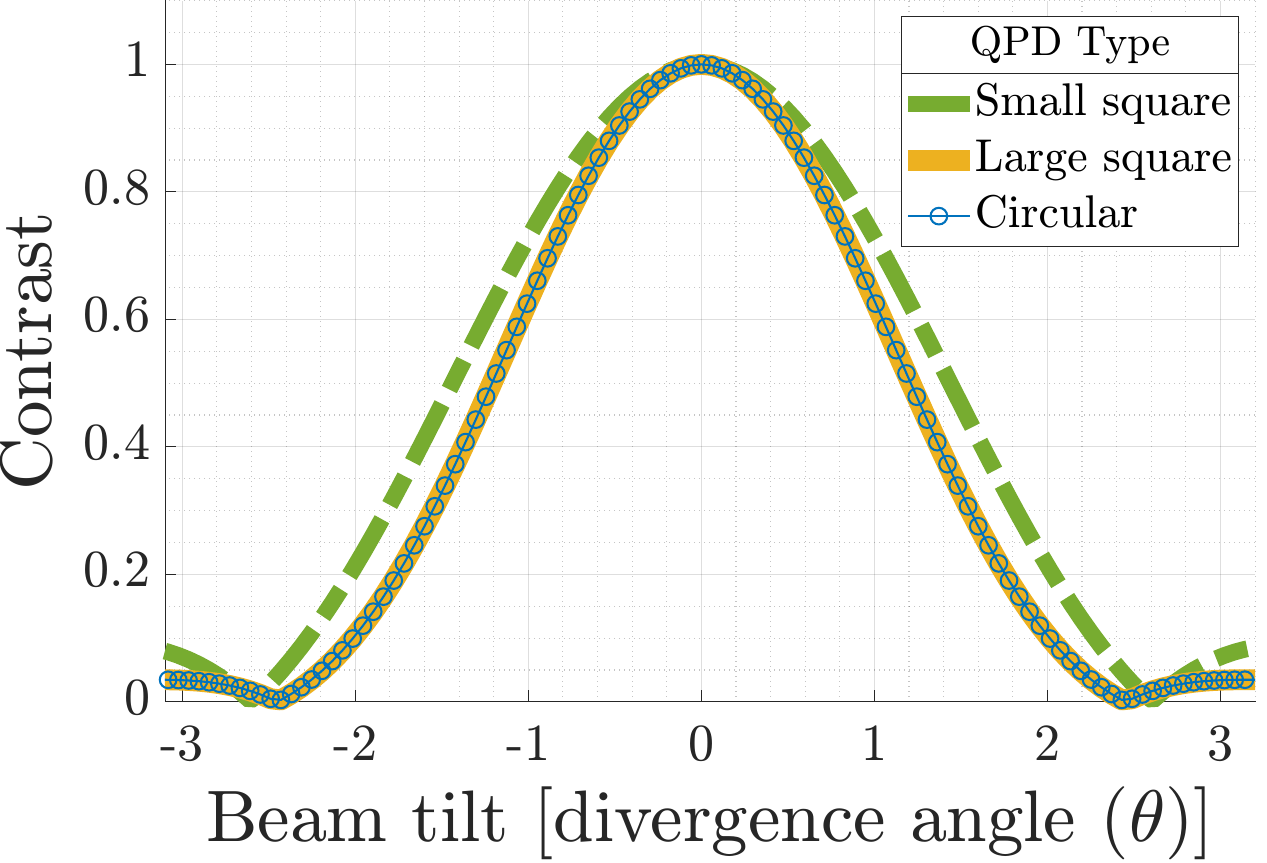}\label{fig:large_angle_contrast}}
   \sidesubfloat[]{\includegraphics[width = .44\textwidth]{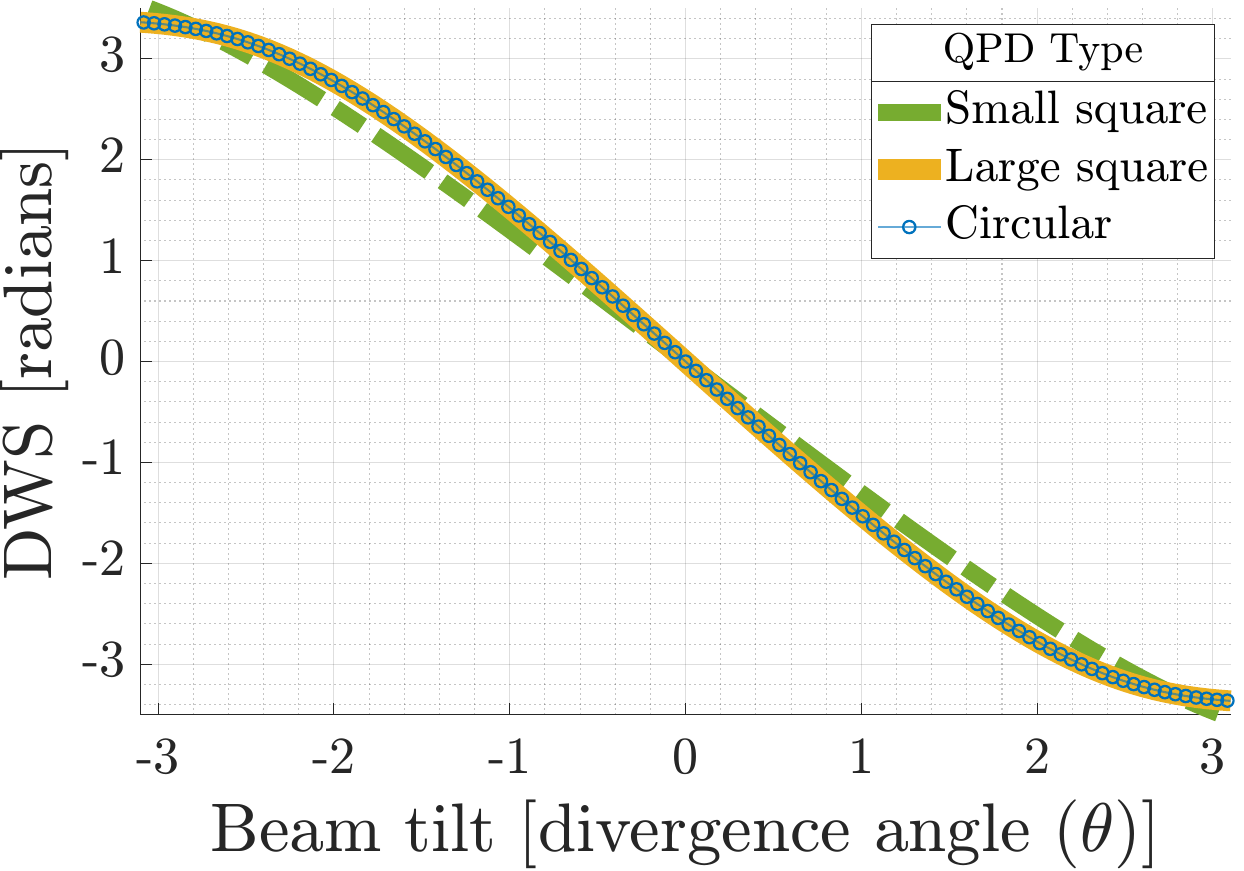}\label{fig:large_angle_DWS}}
    \caption{(a)  Contrast and (b) DWS signals calculations for the three QPD sizes simulated:  $\sqrt{2}/2\,$mm ``small'' square (dashed green), 1 mm ``large'' square (yellow), and 1 mm-diameter circular (blue) \gls{QPD}s. Values on the $x$-axis are in units of divergence angle ($\theta = 840$ $\muup$rad) for tilt of one of the two beams.}
\end{figure}

The range of angles for this rotation is shown in Fig.$~$\ref{fig:large_angle_contrast} in units of the beam divergence angle, $\theta$ (i.e., we plot $\alpha/\theta$), where 
\begin{equation}
    \theta = 
    \arctan\left( \frac{\lambda}{\pi w_0}\right)
    =
    842\,\muup\textrm{rad}.
\end{equation}
From Fig.~\ref{fig:large_angle_DWS} we found for all tested \gls{QPD} shapes and sizes that the \gls{DWS} signal, which is a photodetector phase readout scheme used to track angular alignment, is approximately linear within a beam tilt range of at least $\alpha = \pm 1\cdot\theta$. Within this range, the interferometer has more than 60\% contrast, which is high enough to ensure quality of all readout signals. We rounded the tilt range up to $\pm 900\,\muup$rad and used this for all of our test cases.

\subsubsection{Beam size}\label{sn:beam_size}

The laser itself and the waist sizes of the \gls{MB} and \gls{LO} are part of the ongoing LISA design \cite{laser_progress_2022}. 
The nominal value of $w_0=402\,\muup$m for waist radii was chosen not only as a rough match to the expected mission parameters, but also because it represents the optimal fitting and good overlap of a Gaussian beam with a $900\,\muup$m tophat beam diameter (a value we have used for ongoing simulation of tophat-Gaussian interference at the \gls{ISI}). 
\cref{params_table} shows that we have varied this waist radius for either of the beams at a time by up to $\pm 10\%$, and tested its effect on the \gls{TTL} coupling. 
Physically, change to a single waist size term manifests as imperfect mode matching at the \gls{QPD}.
The $\pm 10\%$ variation was chosen from an order of magnitude estimate of how well the spot sizes can be matched on the optical bench. 
Our results of the waist size variations performed are shown in \cref{sub:w0_diff}.

\subsubsection{Lateral offset of the MB}\label{MB_Intro_Lat}
In this first scenario of misalignment coupled with beam tilt, the \gls{MB} is laterally offset ($x_{\textrm{off}}$) along the $x$-axis before the rotation about the $y$-axis described in \cref{sn:beam_tilt}. 
The value was set to be $0.2\,$mm, which represents one-half of the Gaussian beam radius. 
This magnitude of lateral beam shift is significantly higher than what we expect for LISA. 
However, this range allowed comfortable modeling of the TTL coupling, while maintaining functional contrast (here $c > 0.5$).
This type of lateral misalignment of the \gls{MB} could originate from tolerances during the assembling of the optics prior to the imaging system.
The simulation results are discussed in \cref{sub:x_off}.

\subsubsection{Lateral shift of the MB's CoR}\label{COR_Intro_Lat}

The beam tilts investigated within this paper can originate from either spacecraft or test mass angular jitter. Ideally, the pivot of both these jitters would laterally coincide with the center of the test mass. 
The imaging system then projects this pivot onto the \gls{QPD} center.

We now consider cases where the \gls{MB} does not pivot around the \gls{QPD} center, but around a point laterally shifted by up to $x=1\,$mm.
This could originate from a laterally shifted pivot of the test mass or spacecraft jitter. Alternatively, this could be caused by non-ideal imaging, or by both (a non-ideal pivot location and imperfect imaging) at the same time. 
An offset of $1\,$mm at the \gls{QPD} translates to a displacement of the jitter pivot point by $2.5\,$mm, which we estimate to be a reasonable to optimistic value for spacecraft jitter but an exaggerated case for the test mass.

Please note that in the case of a lateral displacement of the center of rotation of the spacecraft or test mass, additional geometric TTL effects would occur, which we do not consider in this study. 
However, here we focus solely on the TTL due to the described lateral displacement at the \gls{QPD} level.
Our simulation results for this test are shown in \cref{sub:x_cor}.

\subsubsection{Longitudinal offset of the waist}\label{MB_Intro_Long}

Longitudinal offset ($z_\textrm{off}$) repesents a defocus caused by mode mismatch.
This value was varied within a range of $\pm47.6\,$mm, which is approximately $\pm 10\%$ of the interfering beams' Rayleigh ranges, given by
\begin{equation}
    z_\textrm{R}
 = 
    \frac{\pi w_0^2}{\lambda} \;.
\end{equation}
This allows us to investigate the effects of relatively large offsets without great influence from divergence effects.
This also accounts for the experimental challenge of aligning the beam parameters perfectly with each other.
Within a Rayleigh range, the spot radius increases only by $\sqrt{2}$, so we assume that about $10\%$ is a sensible measure of what can be achieved.
The simulation results for this parameter are given in \cref{sub:z_off}.

\subsubsection{Longitudinal shift of the CoR}\label{COR_Long_Rot}

A shift of the \gls{CoR} along the primary optical axis ($\zcor$) displaces the intersection point of the beam with the \gls{QPD} plane (cf. \cref{subfig:sketch_default}). 
Again, the ranges of misalignment were chosen to maintain contrast.
These values were maximally a $\pm 5\,$mm offset in rotational center, representing a shift of the \gls{CoR} to before ($-$) or after ($+$) the \gls{QPD}. This represents an instance where the \gls{CoR} does not align longitudinally with the pupil plane of the imaging system, causing its effective projection to be displaced from the surface of the \gls{QPD}.
This case represents imperfect imaging as the \gls{MB} \gls{CoR} rotates about a particular pivot point along the longitudinal axis. 
For this case, the $\pm 5\,$mm range was chosen such that the \gls{TTL} signal varied enough to allow for a higher order polynomial fit while the interferometer maintains functional contrast.
The simulation results are discussed in \cref{sub:z_cor}

In the next section, we will describe the three simulation methods and how they allow us to quantify the noise resulting from these \gls{TTL} couplings.

%% file: S3_Methods.tex
\section{Methods and scope}\label{Section3}

We conducted simulations using three distinct techniques to model TTL coupling localized to the \gls{TMI}.
This allowed us to cross-check all results obtained within this paper.
Within this section, we introduce the three methods and describe a useful coordinate transformation to represent the beam fields.
Finally, we demonstrate how beat note phases at each \gls{QPD} quadrant are used to formulate \gls{LPS} and \gls{DWS} signals.

\subsection{Simulation techniques}\label{sub:sim_techs}

The three simulation methods that we applied to every simulation can be described as:

\begin{enumerate}
    \item Analytical Method - a generalized analytical solution for modal overlaps.
    \item Expansion Method - a Taylor expansion of the generalized analytical solution in our simulation's perturbation parameters. 
    \item IfoCAD Method - simulations using the \textit{IfoCAD} software library \cite{ifocad}.
\end{enumerate}   
Because it is based on numerically calculating beam overlap integrals, only the \textit{IfoCAD} method could be applied to a circular \gls{QPD} geometry - the other two methods evaluated analytical overlap integrals for square QPDs of two sizes (small and large), as shown in \cref{sn:pd_dim} because these are more tenable analytically.
The overlap $O_j$ of the beam fields, $A_{\textrm{(LO/MB)}}$, on each quadrant active surface, $S_j$, of these \gls{QPD}s is defined as
\begin{equation}\label{eq:overlap}
    O_j\equiv\iint_{S_j} A_\textrm{LO}(A_\textrm{MB})^*\; d^2 S\;  \;\;\;\;\;\;\;\forall j \in \{1,2,3,4\}
    \;.
\end{equation}
See \cite{Weaver:2021wqy} for the explicit form of this overlap integral.
This explicit form is evaluated in the ``Analytical'' and ``Expansion'' methods.


\subsubsection{Analytical method}

In this method, the superposition of two fundamental Gaussian modes is computed and the corresponding overlap is analytically evaluated. 
The tilts and offsets of the beams are represented via a coordinate transformation. 
The generalized coordinate transformation, $(x,y,z) \rightarrow (x',y',z')$, for beam yaw with any of our four misalignment terms of \cref{sub:param_space} results in the following primed coordinate system:
\begin{equation}\label{eq:coord_transformation}
\left[
    \matrix{
x' \cr
y' \cr
z' 
}
\right]
=
\left[
\matrix{
(1- \cos \alpha) x_\textrm{cor} 
+ (x \cos \alpha - x_\textrm{off})
+ (z-z_\textrm{cor}) \sin \alpha \cr
y \cr
(x_\textrm{cor}-x) \sin \alpha
+(1-\cos \alpha)z_\textrm{cor}
+(z \cos \alpha - z_\textrm{off})  
}
\right]
.
\end{equation}
For analytically computing the overlap integral, we followed the method described in \cite{Wanner:14}, but expanded it for square \gls{QPD}s.
Correspondingly, expansions in the longitudinal coordinate $z$ for the Gouy phase, beam waist, and radius of curvature terms are neglected. 
These terms are dominated by relatively large $k^2$ factors (c.f. \cref{eq:Gouy_phase,eq:Rad_curvature,eq:beam_rad}), so only each $x$ coordinate and the $z$ coordinate in the \gls{HG} plane wave term ($e^{-\mathrm{i}kz}$) of Eq.$\;$(\ref{eq:hg}) are transformed.
Misaligned Gaussian beams can then be represented as
\begin{samepage}
\begin{eqnarray}\label{eq:hg_tran}
	\textrm{HG}_{0,0}(x',y',z') \equiv &
	\sqrt{\frac{2}{\pi w(z)}}
		\exp 
		\left(\frac{-\mathrm i
   k(x'^{2}+y^{2})}{2R(z)}-
		\frac{x'^{2}+y^{2}}{w^2 (z)}\right) 
		\nonumber \\ & \times 
  \exp \left(
  \mathrm i\psi(z) - 
  \mathrm i k z'		
		\right)
		\;.
\end{eqnarray}
\end{samepage}

In our simulations, the overlap integral (cf. \cref{eq:overlap}) was calculated between this misaligned Gaussian and an on-axis Gaussian, representing the \gls{MB} and \gls{LO} respectively.
For the mathematical formulae behind the Analytical Method, see Appendix C.18 of \cite{Weaver:2021wqy}.

\subsubsection{Expansion method}\label{sub:expansion_method}

This method is an approximation of the analytic method. Here, rather than working with exact coordinate transformations, we derive a corresponding model consisting of a superposition of higher-order \gls{HG} modes. 
This follows the well-known principle that any sufficiently paraxial beam can be written as a linear combination of \gls{HG} modes (for a review, see \cite{siegman}). It has the alternative properties that an ``off-axis'' Gaussian measurement beam can be approximated as a sum of ``on-axis'' HG modes. This results in overlap integrals that are easier to evaluate. 
The modal coefficients were derived using identities of the Hermite polynomials  (similar to the work in \cite{Bayer-Helms:84}) and Taylor expansions to seventh order in the misalignment parameters.

However, the order cut-off highlights limitations in this method while also providing good accuracy for cross-checking with other simulation methods. 

In the following, we show two exemplary models derived with the expansion method. For these, we use Hermite polynomials~$\mathrm H$ with indices $m$ and $n$, and the \gls{HG} modes expressed as
\begin{samepage}
\begin{eqnarray}\label{eq:hg}
	\textrm{HG}_{m,n}(x,y,z) =&
	\frac{		\textrm{H}_{m} \Big(\frac{\sqrt{2}x}{w(z)} \Big)
		\textrm{H}_{n} \Big(\frac{\sqrt{2}y}{w(z)} \Big)}{\left({2^{m+n-1}\pi m!n!w(z)}\right)^{1/2}}
		\nonumber \\ & \times 
		\exp 
		\left(\frac{-\mathrm i
   k(x^{2}+y^{2})}{2R(z)}-
		\frac{x^{2}+y^{2}}{w^2 (z)}\right) 
		\nonumber \\ & \times 
  \exp \left(
  \mathrm i (m+n+1)\psi(z) - 
  \mathrm i k z		
		\right)
		\;,
\end{eqnarray}
\end{samepage}
where, for wavenumber ${ k}=\frac{2\mathrm \pi}{\lambda}$ (and wavelength $\lambda=1064\;$nm), the Gouy phase is

\begin{equation}\label{eq:Gouy_phase}
\psi(z)
 = \arctan\left( \frac{2(z-z_0)}{k w_0^2} \right) 
 \;,
\end{equation}
radius of curvature is

\begin{equation}\label{eq:Rad_curvature}
R(z)
 = z - z_0 + 
 \frac{ k^2 w_0^4}{4(z-z_0)}
 \;,
\end{equation}
beam radius is

\begin{equation}\label{eq:beam_rad}
w(z)
 = w_0 \left[{1 + \left( \frac{2(z-z_0)}{k w_0^2} \right)^2}\right]^{1/2}
 \;,
\end{equation}             
and $w_0$ is the waist located at $z=0$.

The first and well-known example considers a simple case of no propagation, where $z=0$. A Gaussian beam with small lateral offset $x_\textrm{off}$ is approximated to first order as a sum of $\textrm{HG}_{0,0}$ and $\textrm{HG}_{1,0}$ modes along with a scaling coefficient,
\begin{eqnarray}
     \nonumber \textrm{HG}_{0,0}(x-x_\textrm{off},y,0)
     & \approx
    \textrm{HG}_{0,0}(x,y,0)+x\left(\frac{2x_\textrm{off}}{w_0^2}\right) \textrm{HG}_{0,0}(x,y,0)
    \\ & \approx \label{eq:exp_example}
    \textrm{HG}_{0,0}(x,y,0)+\frac{x_\textrm{off}}{w_0} \textrm{HG}_{1,0}(x,y,0)
    \;.
\end{eqnarray}
In the second line, the coefficient for $\textrm{HG}_{0,0}$ is unity, and the $\textrm{HG}_{1,0}$ coefficient was derived such that it has no coordinate dependence (this dependence was incorporated into the mode).

The second example demonstrates instead the more complex case of a lateral offset of the beam axis ($\xoff$) coupled to beam tilt ($\alpha$).
Applying the Expansion Method to first order in both terms (offset and tilt), the multivariate expansion of the misaligned Gaussian is overall a second order solution: 
\begin{eqnarray}\label{eq:order1_exp}
    \textrm {HG}_{\textrm{0,0}} (x',y',z') \approx & 
    \left\{\textrm {HG}_{\textrm{0,0}} (x,y,z) 
        \times
        \left[
            1 
            - \textrm{i} k \xoff \alpha 
            + \textrm{i} k \frac{ \xoff \alpha   z}{R(z)}
            + 2 \frac{ \xoff \alpha z}{w^2(z) }
        \right]
    \right\}
        \nonumber\\&
        +
    \left\{
            \textrm{HG}_{\textrm{1,0}} (x,y,z) \times
            \left[
                \frac{w_0}{2} 
                \left(
                    1-i \frac{z}{z_{\textrm R}} 
                \right)
            \right]
    \right.
    \nonumber\\&
    \left.
            \times 
            \left[
                2 \frac{ \xoff }{w^2(z)}  
                - \textrm{i} k \frac{ \alpha z}{R(z) } 
                - \frac{2 \alpha z}{w^2(z)}  
                + \textrm{i} k \alpha   
                + \textrm{i} k \frac{\xoff   }{R(z)}  
            \right]
    \right\}
    \nonumber \\& 
    +
     \frac{w_0^2}{4}
    \left\{
        \left[
            \textrm {HG}_{\textrm{0,0}} (x,y,z)
            \times
            \left(1+\left( \frac{z}{z_{\textrm R}}\right)^2\right)
            \right.
            \right.
            \nonumber\\&
            \left.
            \left.
            +
            \textrm {HG}_{\textrm{2,0}} (x,y,z) 
            \times
            \sqrt{2}\left(1- i \frac{z}{z_{\textrm R}}\right)^2
        \right]
    \right.
    \nonumber\\&
    \times
    \left.
        \left[
        2 \textrm{i} k\frac{ \xoff \alpha  }{w^2(z)} 
        - k^2 \frac{ \xoff \alpha  }{R(z)} 
        + k^2 \frac{\xoff \alpha  z}{R^2(z)} 
        \right.
    \right.
    \nonumber\\&
    \left.
        \left.
        - 4\frac{ \xoff \alpha z}{w^4(z)} 
        - 4\textrm{i} k \frac{ \xoff \alpha    z}{R(z) w^2(z)}
        \right]
    \right\}
    \, .
\end{eqnarray}
We show here only the second order result, rather than the seventh order result used in our simulations, due to the length of the expansion.


\subsubsection{IfoCAD method}

The third method uses \textit{IfoCAD}, a C++ software library developed at the Albert Einstein Institute (AEI) for modelling and parameterizing three-dimensional interferometric setups \cite{ifocad}. 
In \textit{IfoCAD}, we built the simplified interferometric setup as described in \cref{sub:nom_setup} with the two interfering beams and the diode (\gls{QPD}).
\textit{IfoCAD} traces the beams to the diode using dedicated routines, computes the overlap integral of the two beams and further evaluates signals using the methods described in \cite{wanner2012methods}. 
The \textit{IfoCAD} framework allows for calculating various signals on a \gls{QPD}, including \gls{DWS} and different \gls{LPS} types (such as \gls{LPSA} and \gls{LPSC}).  

For our simulations of the \gls{TMI}, two circular Gaussian beams were defined. We then simulated the different test cases (variations of beam properties, the center of rotation, or the QPD location) by tuning the corresponding parameters in \textit{IfoCAD}. Afterwards, the \gls{MB} was tilted and both beams were propagated to the photodiode. The overlap of this rotated beam with the nominally aligned beam (\gls{LO}) was numerically calculated using \cref{eq:overlap}, and \gls{LPSA} and \gls{LPSC} signals were computed.


\subsection{Quadrant phases \& geometric and non-geometric TTL}\label{sub:beatnote_signal_phase}

All three simulation methods calculate the overlap integral (\cref{eq:overlap}) for each quadrant.  
From this, the corresponding phase is computed using the well-known in-phase ($I$) and  quadrature ($Q$) demodulation, described for instance in \cite{WANNER20124831}. 
It follows from demodulation that the phase over each quadrant, $\phi_j$, results from the imaginary and real parts of the overlap integral $O_j$, of \cref{eq:overlap},
\begin{equation}\label{eq:signal_phase}
    \phi_j = \arg \left[ O_j \right]
    = \arctan \left(\frac {\mathrm{Im} [O_j]}{\mathrm{Re}  [O_j] } \right)
    \;.
\end{equation} 
To reduce numerical errors, we perform operations described in \cite{WANNER20124831} and separate the plane wave phase dependence of each beam's central ray from the more slowly varying residual wavefront profile $A_\mathrm{(LO/MB)}'$, using the wavenumber for each beam $k_\mathrm{(LO/MB)}$,
and the total propagation distance of each beam $D_\mathrm{(LO/MB)}$:
\begin{equation}\label{eq:beam_prof}
    A_\mathrm{(LO/MB)} \rightarrow A_\mathrm{(LO/MB)}' e^{- \mathrm i k_\mathrm{(LO/MB)} D_\mathrm{(LO/MB)}}
    \;.
\end{equation}
As shown in \cite{Hartig_2023}, the phase at each $j$-th QPD quadrant can then be written as 
\begin{equation}\label{eq:TTL_breakdown}
    \phi_j = 
    \overbrace{
    \left[
    k_\textrm{MB} D_\textrm{MB} - 
    k_\textrm{LO} D_\textrm{LO}
    \right]}^{\textrm{Geometric\,TTL\,Phase\,} \Phi_{\textrm{GEO/OPD}}}
    +
    \underbrace{\arg \left [
    \iint_{S_j}  
    \left[
    A_\textrm{LO}' (A_\textrm{MB}')^*
    \right]
    d^2S
    \right ]}_{\textrm{Non-Geometric\,TTL}}
    \;.
\end{equation}

Note that in \cref{eq:TTL_breakdown} we have split the phase into a sum of two \gls{TTL} components: ``geometric'' and ``non-geometric.'' 
Geometric \gls{TTL} is the phase change caused by a change in the \gls{OPD}. 
This component can be described purely by geometry or simple trigonometric relations \cite{Hartig_2022}. 
The non-geometric TTL term refers to additional coupling mechanisms which originate from wavefront and detector properties \cite{Hartig_2023,Sasso_2018,Weaver:20}. 
The non-geometric terms also depend on the precise definition chosen for the \gls{LPS}.

\subsection{Interferometric signal calculation}
\label{sub:signals_calc}

\begin{figure}[!htbp]
    \includegraphics[width=.3\textwidth]{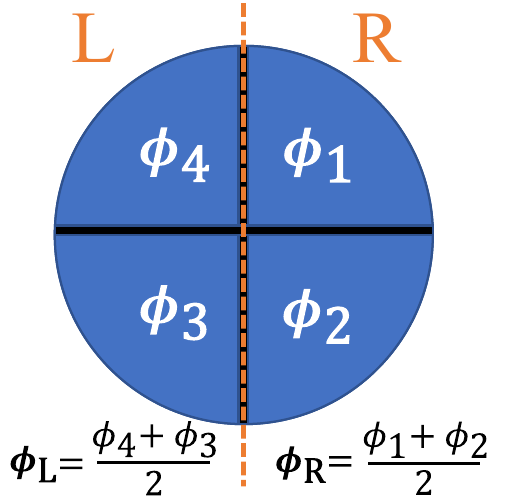}
    \caption{The phase terms at
    each photodiode quadrant of a circular QPD are labeled in the diagram, bisected for left ``L'' and right ``R'' as in the DWS description.}
    \label{fig:phase_qpd}
\end{figure}

In this study, we focus on three specific signals. The first is the \gls{DWS} scheme used to track angular alignment. 
Throughout this paper, we investigate only rotations about the $y$-axis. This causes a horizontal beam tilt that is measured by the phase difference of the right and left halves of the \gls{QPD}.
The corresponding \gls{DWS} signal is defined by
\begin{equation}
\label{equation:dws}
\mathrm{DWS} \equiv
 		 	\left[ 
 		 	\phi_\textrm{R} - 
     \phi_\textrm{L} 
 		 	\right]
     \; .
\end{equation}
The \gls{QPD} diagram in \cref{fig:phase_qpd} may be used as a reference for the quadrant indices.

Furthermore, we consider the longitudinal pathlength signals used to track displacements. 
For these, we focus on the \gls{LPSA} and the \gls{LPSC} formulations. \gls{LPSA} represents the ``averaged phase,'' i.e., the arithmetic mean of quadrant phases, 

\begin{equation}
\label{equation:lpsa}
\mathrm{LPSA} \equiv
 		\frac{1}{4k}
 		    \sum_{j=1}^4 \arg 
 		    \left[ 
 		    O_j 
 		    \right]
       =
 		\frac{1}{4k}
 		    \sum_{j=1}^4 
       \phi_j
 		 		.
\end{equation}
Finally, \gls{LPSC} is the formulation which was employed in LISA Pathfinder (LPF).
It is defined as the argument of the sum of all complex amplitudes, converted to lengths by $1/k$,
\begin{equation}
\label{equation:lpsc}
\mathrm{LPSC} \equiv
 	\frac{1}{k}
 		 \arg \left[\sum_{j=1}^4 O_j \right]
    	 	.
\end{equation}
Now, consider that in \cref{eq:TTL_breakdown} the geometric TTL phase clearly has no quadrant dependence, so this geometric TTL phase is canceled in the DWS formulation.
More importantly, however, geometric \gls{TTL} contributes equally to both LPS formulations.
This motivates us to work with a nominal setup that has zero TTL coupling and simulate then cases of primarily non-geometric \gls{TTL} introduced by our misalignment parameters.
Thereby, we focus on scenarios where the non-geometric TTL is predominant, leading to differences between LPSA and LPSC. 
This is true except for one case, where we show that geometric \gls{TTL} dominates and LPSA and LPSC perform equally, as expected.

The definitions we have provided for the \gls{LPS} and \gls{DWS} signal in this section represent those of the conventional signal extraction architecture \cite{WANNER20124831, wanner_readouts}, where each of the four quadrants of the \gls{QPD} are tracked separately. 
The architecture currently proposed for \gls{LISA}, described in \cite{PhysRevApplied.14.054013}, would instead employ a different set of four tracking loops operating directly on the length and angles. 
Each loop can be individually optimized for the dynamic range and noise characteristics of that particular observable. 
This allows for an improvement in signal-to-noise ratio and robustness to cycle slips, and it is especially useful at the \gls{ISI} where low received beam powers are expected. 
It was likewise shown in \cite{PhysRevApplied.14.054013} that for sufficiently high loop gains, the new architecture is equivalent to the \gls{LPSA}. 
Therefore we consider here that the conventional definition remains appropriate for simulation of the \gls{TMI}, and we thus also compare the performance of the phase signal of the new architecture under the assumption of a high-loop gain (\gls{LPSA}) with the performance of one alternative signal definition (\gls{LPSC}). 
In the next section, we report the results of our study in terms of these signals.

%% file: S4_Results.tex
\section{Simulation results}\label{Section4}

In this section, we present simulation results for the two \gls{LPS} candidates: \gls{LPSC} and \gls{LPSA}. 

\subsection{Results for variations in the waist sizes of the beams}\label{sub:w0_diff}
\begin{figure}[!htbp]
\centering
\sidesubfloat[]
{\includegraphics*[trim = 5 0 0 5, width= 0.45\textwidth]{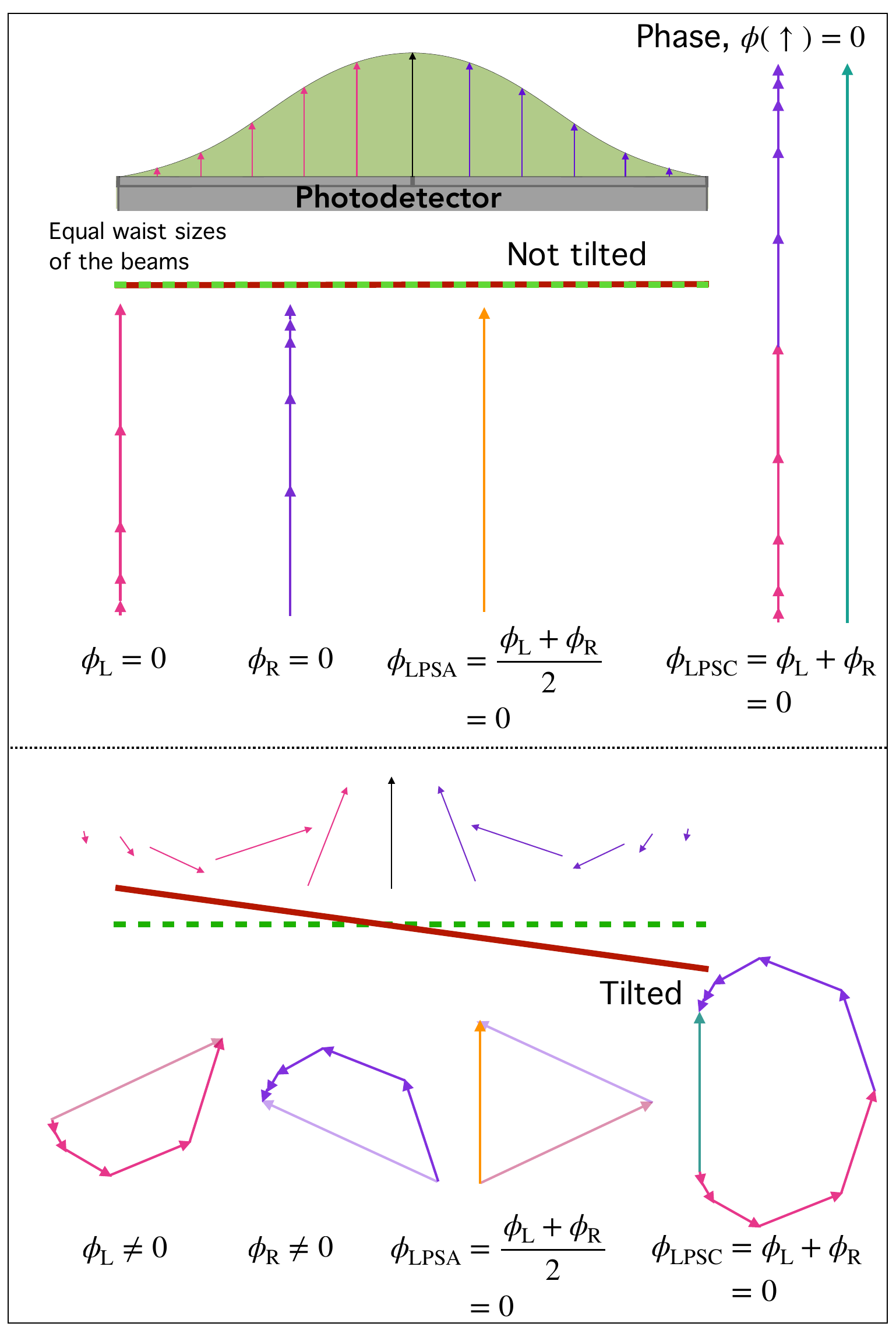} \label{sub:w0_diffa}}
\sidesubfloat[]
{\includegraphics*[trim = 5 0 0 5, width= 0.45\textwidth]{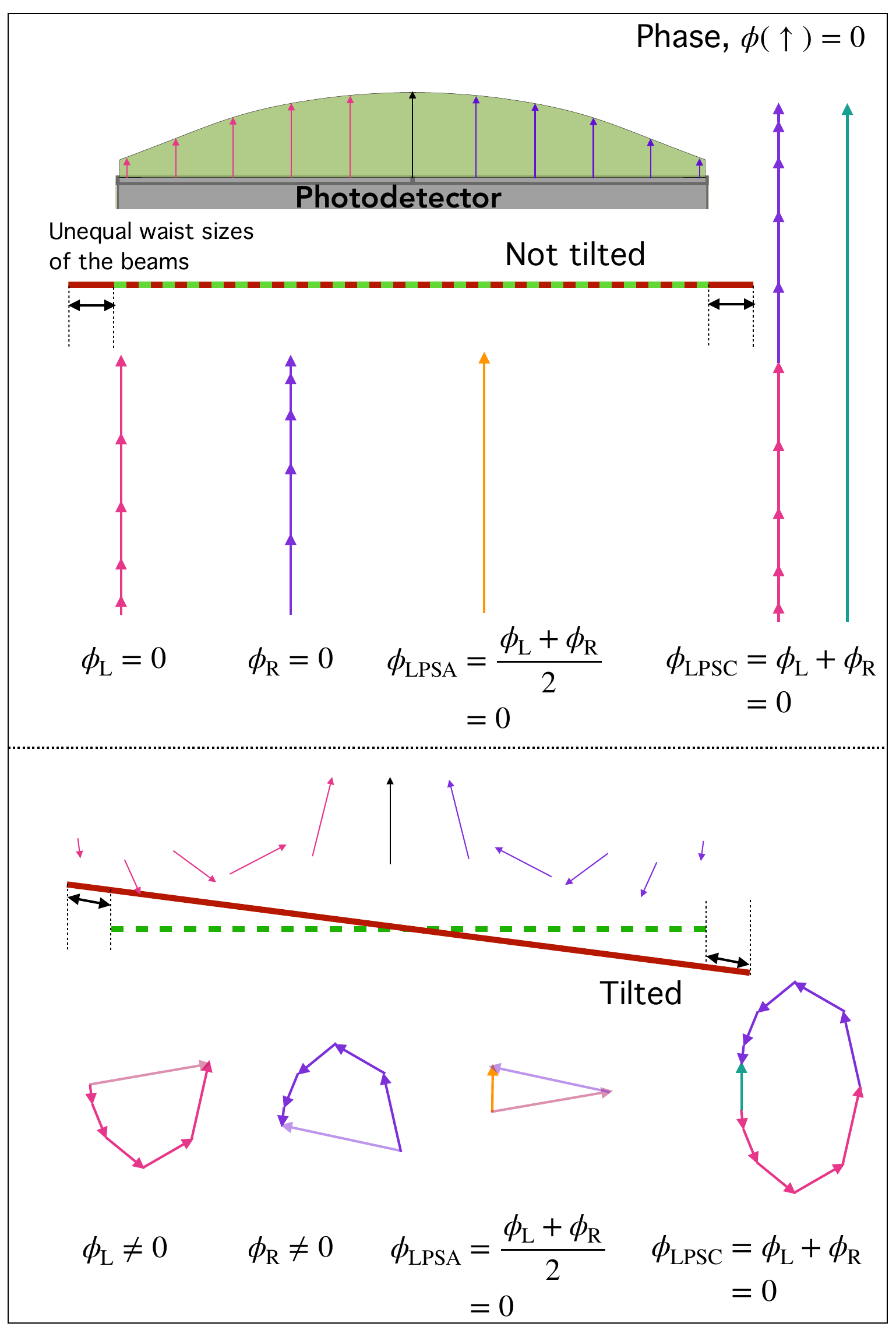} \label{sub:w0_diffb}}

\caption{\label{fig:wavefront_overlap}  Graphical illustration of vanishing non-geometric TTL coupling in two cases with beam rotation around the centre of a QPD. Image (a) shows the case of matched waist sizes, image (b) a case where the \gls{MB} (red flat wavefront), has a larger waist than the \gls{LO} (green-dashed flat wavefront). 
The total \gls{LPSA} ($1/k$ times the angle of the orange vector) and \gls{LPSC} ($1/k$ times the angle of the green vector) signals are estimated via vector sums of the individual phasors. 
See text and \cite{Hartig_2023} for more information. 
}
\end{figure}
We conducted simulations which varied waist sizes of the \gls{MB} and \gls{LO} separately.
Here, we varied the waist radii by $\pm 10\,\%$ in steps of $5\,\%$ around the nominal value of $\SI{0.402}{mm}$.
This resulted in a negligible TTL coupling, due only to numerical error ($\mathcal{O}$ $\sim \SI{e-17}{m}$ to $\SI{e-23}{m}$) for all simulation methods and geometries. In this scenario, geometric TTL coupling plays no role due to the absence of differences in the optical pathlengths of the beams. 
Since the total \gls{TTL} coupling is negligible, we can deduce that the non-geometric \gls{TTL} is likewise negligible in this specific case. We explain this graphically in \cref{fig:wavefront_overlap}.

In the two scenarios of \cref{sub:w0_diffa}, the \gls{LO} and \gls{MB} are Gaussian beams of equal waist sizes, while in \cref{sub:w0_diffb} the waist sizes are unequal.
In the top of each figure, we sketch a rough estimate of the amplitude resulting from the superposition of the two beams at the photodetector.
This amplitude shows slight variation as the \gls{MB} waist size varies and serves for defining the lengths of the vectors (phasors) in the lower parts of the images.
Below these amplitude diagrams in each subfigure, we trace the wavefronts of the two beams in red (\gls{MB}) and green (\gls{LO}) and allow for clipping when the \gls{MB} size is varied in \cref{sub:w0_diffb}. 
Each of these wavefronts is flat, due to the waists being located on the photodiode (cf.~\cref{Sec:SimSetup,fig:IS_wavefront}).
The amount by which the red wavefront is above or below the green wavefront defines the angle of the corresponding phasors.
The bottom half of each subfigure depicts these same wavefronts respectively for the cases in which the \gls{MB} is tilted with respect to the \gls{LO}.

For all four scenarios, we explicitly show the sum of the phasors reduces to zero, signifying that the non-geometric \gls{TTL} is negligible for the case of waist size variation due to this phase antisymmetry.
Please note that the $\phi_\text{LPSA}$ was illustrated via a vector sum. 
This is not the usual procedure, but possible in this specific case due to the equal lengths of the phasors illustrating the complex amplitudes on the left (light red phasor) and right (light purple phasor) halves of the photodiode. 
This implies that for such cases with equal amplitude on the left and right halves, the \gls{LPSA} and \gls{LPSC} are identical.
Please see \cite{Hartig_2023} for a detailed explanation of this type of illustration. 
%

\subsection{Results for lateral offset of beam axes}\label{sub:x_off}

\begin{figure}[!htbp]
\begin{center}
    \includegraphics[width=.35\textwidth,
    trim=0 .2cm 0 0, clip]{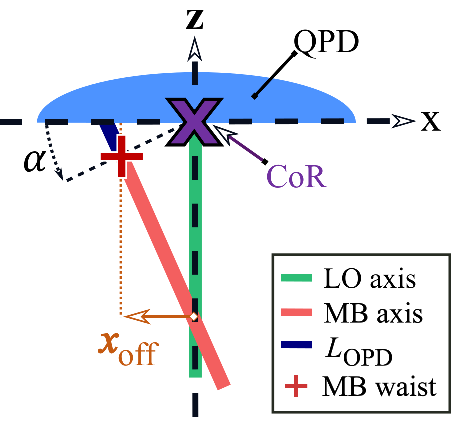}
  \end{center}
  \caption{\label{fig:sketch_lps_x_off}Test geometry where the measurement beam's (MB) center of rotation remains the \gls{QPD} center (shown by a purple ``X''), but the beam axis is laterally offset by a distance $x_\textrm{off}$ before rotating in $\alpha$. These rotations yield an additional pathlength $L_\textrm{OPD}$ extending from the red ``+'' which marks the new \gls{MB} waist to the intersection point of the beam axis with the photodiode. The depiction is not to scale, as even the largest offset of 0.2\,mm still allows beam overlap.}
\end{figure}

In this simulation, we tested the \gls{TTL} coupling due to a lateral offset ($\xoff$) of the \gls{MB} with respect to the \gls{QPD} center.
This is shown in \cref{fig:sketch_lps_x_off} and includes an optical pathlength difference $\lopdeq$ (as a blue segment of the \gls{MB}), which for this scenario is
\begin{equation}
    \lopdeq(\alpha,x_\textrm{off,MB})= -x_\textrm{off,MB} \tan (\alpha) \approx -x_\textrm{off,MB} \alpha
    \,. \label{eq:LOPD-case-2}
\end{equation} 
%

The optical pathlength difference $\lopdeq$ defined in \cref{eq:LOPD-case-2} was derived from the difference between the propagation distance of the tilted \gls{MB} and the propagated distance of the nominal \gls{MB}. 
However, this equals the definition given in \cref{{eq:TTL_breakdown}}. To see this, one can set $k_{\textrm{MB}}=k_{\textrm{LO}}$ and neglect the non-geometric elements in this equation, to find
\begin{equation}
    L_{\textrm{OPD}} (\alpha) \Rightarrow 
    \left[ 
    D_{\mathrm{MB}}(\alpha)-D_{\mathrm{LO}}(\alpha) 
    \right]
    -
        \left[ 
    D_{\mathrm{MB}}(0)-D_{\mathrm{LO}}(0) 
    \right]\;.
\end{equation}
The propagation distances for the \gls{LO} cancel from this equation, because the \gls{LO} does not rotate. This results in the described comparison of the \gls{MB} propagation distances.

We also quantify non-geometric contributions in each test via a functional fitting of the \gls{LPS} to its misalignment. 
Calling each misalignment parameter $\chi$, the $\lopdeq$ and non-geometric contributions combine to induce a change to the \gls{LPS}, $\Delta \mathrm{LPS}$, for which we provide a functional fitting of the form
\begin{equation}\label{eq:fun_fit}
    \Delta \textrm{LPS}(\alpha, \chi) \approx 
    \lopdeq + \sum_{i,j=1}^4 \left(c_{i,j} \alpha^i \chi^j \right) \;.
\end{equation}
Here, $c_{i,j}$ is the scaling coefficient for a tilt, $\alpha$, of order $i$ coupled to a particular misalignment, $\chi$, of order $j$.
These we accurately truncate to fourth order (i.e., a cutoff of $i,j=4$).
The $c_{i,j}$ non-zero values therefore represent the strongest coupling coefficients in a polynomial expansion of non-geometric \gls{TTL} obtained via least squares-fit in terms of beam tilt, $\alpha$. 
Square \gls{QPD} coefficients in this fitting were derived from data of the Analytical Method, while those for the circular \gls{QPD} were from \textit{IfoCAD} data.
These two methods are sufficient to generate third-order coefficients.
For instance, the functional form for this \gls{MB} tilt-misalignment coupling (where $\chi = x_\textrm{off,MB}$) is
\begin{equation}
    \Delta \textrm{LPS}(\alpha,x_\textrm{off,MB}) \approx 
    \lopdeq+\alpha(c_{1,1} x_\textrm{off,MB}+c_{1,3} x_\textrm{off,MB}^3) \;.
\end{equation}
Values calculated from fitting to this coupling are shown in \cref{table2}.

\begin{table}[!htbp]
\caption{Fit coefficients for lateral offsets of the \gls{MB}'s beam axis ($x_\textrm{off,MB}$).}\label{table2}
\scaletable{
\begin{tabular}[h!]{c c c c }
\br
    Coefficient &  Small Square QPD Range & Circular QPD Range & Large Square QPD Range\\
    \br
   LPSC $c_{1,1}$ $\left(\frac{\mathrm{pm}}{(\mathrm{mm})(\muup\mathrm{rad})}\right)$  & $(6.450 \pm 0.017) \times 10^{2}$ & $(5.544 \pm 0.013) \times 10^{2}$ & $(5.247 \pm 0.016) \times 10^{2}$
   \\
LPSA $c_{1,1}$ $\left(\frac{\mathrm{pm}}{(\mathrm{mm})(\muup\mathrm{rad})}\right)$  & $(8.996 \pm 0.002) \times 10^{2}$ & $(8.604 \pm 0.001) \times 10^{2}$ & $(8.488) \times 10^{2}$
\\
\mr
LPSC $c_{1,3}$ $\left(\frac{\mathrm{pm}}{(\mathrm{mm}^3)(\muup\mathrm{rad})}\right)$  & $(2.150 \pm 0.046) \times 10^{2}$ & $(2.108 \pm 0.034) \times 10^{2}$ & $(2.211 \pm 0.036) \times 10^{2}$
   \\
LPSA $c_{1,3}$ $\left(\frac{\mathrm{pm}}{(\mathrm{mm}^3)(\muup\mathrm{rad})}\right)$  & $(1.637 \pm 0.008) \times 10^{1}$  & $(6.651 \pm 0.284)$ & $(5.850 \pm 0.500)$
\\
\br
\end{tabular}
}
\end{table}

\begin{figure}[!htbp]
\centering
    \sidesubfloat[]{\includegraphics[width = .9\textwidth,
    trim=.05cm .05cm .05cm 0cm, clip]{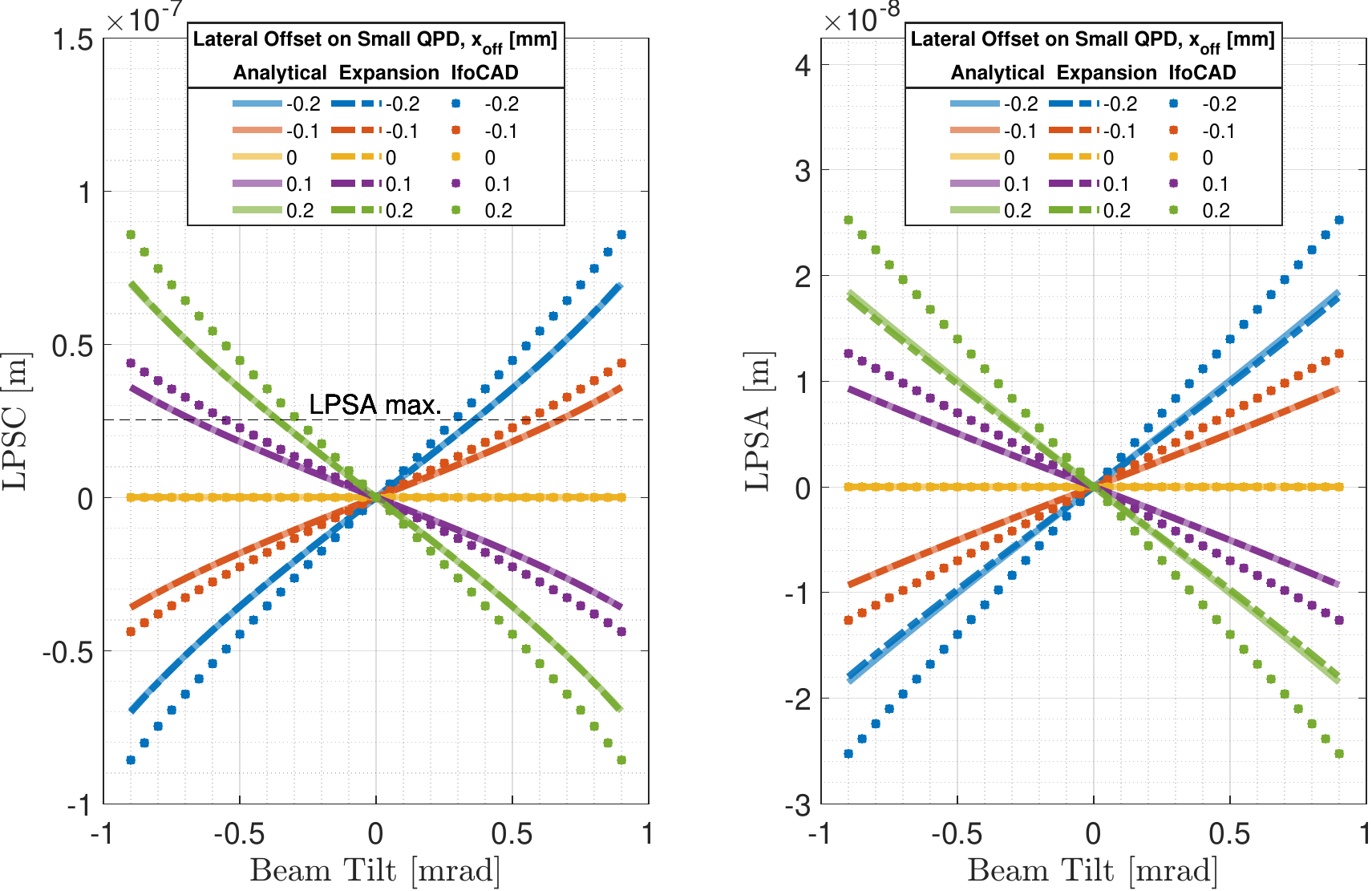}\label{fig:rt2_x_off}}
    \\
    \vspace{.5cm}
   \sidesubfloat[]{\includegraphics[width = .9\textwidth,
    trim=.05cm .05cm .05cm 0cm, clip]{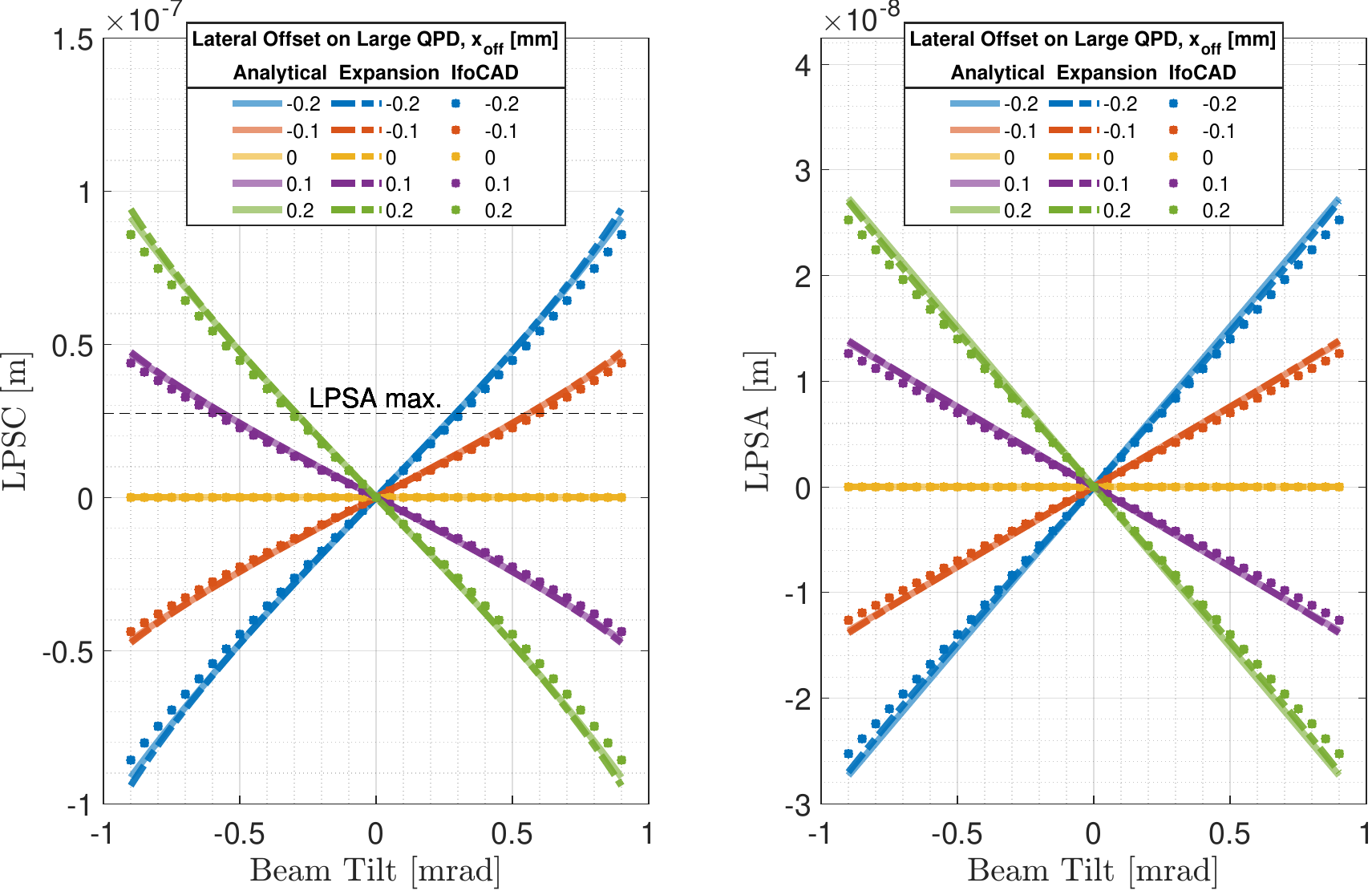}\label{fig:1mm_x_off}}
   \\
    \caption{\label{fig:full_x_off} 
    TTL coupling in the case of lateral shifts of the measurement beam.
    Shown are the LPSC  (left figures) and LPSA (right figures) computed with the analytic and expansion method for ``small'' square QPDs (upper figures, diameter $1/\sqrt2\,$mm) and ``large'' square QPDs (lower figures, $1\,$mm diameter) in comparison to the results for $1\,$mm circular QPDs implemented in IfoCAD.
    The dashed line, ``\gls{LPSA} max.'', denotes the maximal \gls{LPSA} value in each row of plots for comparison of magnitudes in the two formulations. }
\end{figure}

Plots of \gls{LPSC} and \gls{LPSA} calculated by the three simulation methods are shown in \cref{fig:full_x_off} for the ``small"  $\sqrt{2}/2$\,mm-width and ``large" 1\,mm-width square \gls{QPD}s, respectively, against the \textit{IfoCAD} Method 1\,mm-diameter circular \gls{QPD}.
These show that the LPSA is less susceptible to TTL coupling due to lateral offsets of the \gls{MB}'s axis, as the \gls{LPSC} magnitude is nearly three times greater.
These results also demonstrate the influence of \gls{QPD} geometry on pathlength noise, particularly, the discrepancy between the circular and square \gls{QPD} geometries.
While the plots emphasize differences for interferometric simulation with the two square \gls{QPD} sizes, they also display reasonable agreement ($\pm100\,\muup$rad) for the 1\,mm circular and large square \gls{QPD}s (which more closely mimic the circular \gls{QPD}). 
The square \gls{QPD} implementation in \textit{IfoCAD} was verified to agree well with the other two methods (cf. \ref{app:square_qpd}).

\begin{figure}[!htbp]
\begin{center}
    \includegraphics[width=.35\textwidth]{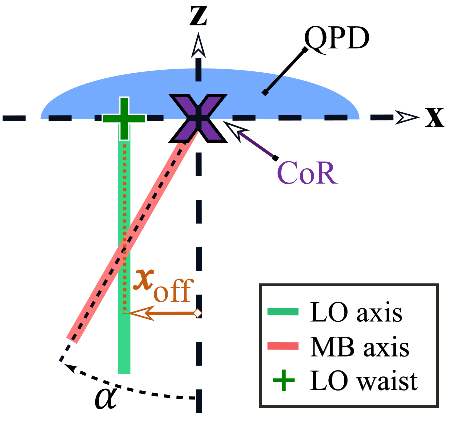}
  \end{center}
  \caption{\label{fig:sketch_lps_LO_x_off}Test geometry where measurement beam center of rotation remains the \gls{QPD} center (shown by a purple ``X''), but the \gls{LO} waist is laterally offset a distance $x_\textrm{off}$. The green ``+'' marks the new \gls{LO} waist location at the photodiode.}
\end{figure}

Now, consider the effect of offsets to the \gls{LO}, rather than the \gls{MB}, while still applying a rotation to the \gls{MB} axis.
When we laterally offset the \gls{LO} instead of the \gls{MB} by $\xoff$, the new scenario is as depicted in \cref{fig:sketch_lps_LO_x_off}.
%
%
For this lateral offset of the \gls{LO} instead of the \gls{MB}, the pathlength signals look nearly identical to the plots of \cref{fig:full_x_off}. 
We have verified this equality to four significant digits beyond the \gls{LPS} ranges shown in \cref{fig:1mm_x_off}.
Although the total couplings agree at this precision, the first-order coefficients in \cref{table3} for non-geometric \gls{TTL} fittings differ from the \gls{MB} lateral offset case in \cref{table2}. 
In changing from an \gls{MB} offset to an \gls{LO} offset, the geometric \gls{TTL} goes to zero so that the functional form for the total TTL coupling in this case is
\begin{equation}
    \Delta \textrm{LPS}(\alpha,x_\textrm{off,LO}) \approx 
    \alpha(c_{1,1} x_\textrm{off,LO} +c_{1,3} x_\textrm{off,LO}^3) \;.
\end{equation}
However, since this total TTL coupling is the same as for the shifted \gls{MB}-axis, the missing geometric TTL-term here must be nearly 
compensated by an additional non-geometric TTL contribution.
Small differences between the two setups of \cref{fig:sketch_lps_x_off,fig:sketch_lps_LO_x_off} are expected due to asymmetrical clipping across the \gls{QPD} inactive regions.

\begin{table}
\caption{Fit coefficients for lateral offset of the LO ($x_\textrm{off,LO}$).}\label{table3}
\scaletable{
\begin{tabular}[h]{c c c c }
\br
    Coefficient &  Small Square QPD Range & Circular QPD Range & Large Square QPD Range\\
    \br
   LPSC $c_{1,1}$ $\left(\frac{\mathrm{pm}}{(\mathrm{mm})(\muup\mathrm{rad})}\right)$  & $(-3.550 \pm 0.017) \times 10^{2}$ & $(-4.457 \pm 0.013) \times 10^{2}$ & $(-4.754 \pm 0.017) \times 10^{2}$
   \\
LPSA $c_{1,1}$ $\left(\frac{\mathrm{pm}}{(\mathrm{mm})(\muup\mathrm{rad})}\right)$  & $(-1.004 \pm 0.002) \times 10^{2}$ & $(-1.397 \pm 0.001) \times 10^{2}$ & $(-1.512 \pm 0.001) \times 10^{2}$
\\
\mr
LPSC $c_{1,3}$ $\left(\frac{\mathrm{pm}}{(\mathrm{mm}^3)(\muup\mathrm{rad})}\right)$  & $(2.150 \pm 0.046) \times 10^{2}$ & $(2.108 \pm 0.034) \times 10^{2}$ & $(2.211 \pm 0.036) \times 10^{2}$
   \\
LPSA $c_{1,3}$ $\left(\frac{\mathrm{pm}}{(\mathrm{mm}^3)(\muup\mathrm{rad})}\right)$  & $(1.637 \pm 0.008) \times 10^{1}$  & $(6.651 \pm 0.284)$ & $(5.850 \pm 0.500)$
\\
\br
\end{tabular}
}
\end{table}

Please note that it might appear from the coupling coefficients that the TTL effect in the \gls{LPSC} formulation is below that of the \gls{LPSA} formulation. 
However, the cancellation between geometric and non-geometric TTL results here in lower TTL noise for \gls{LPSA}  than \gls{LPSC}.

\subsection{Results for lateral offset of measurement beam center of rotation} \label{sub:x_cor}

\begin{figure}[!htbp]
\begin{center}
\includegraphics[width=.35\textwidth,
    trim=0 .1cm 0 0, clip]{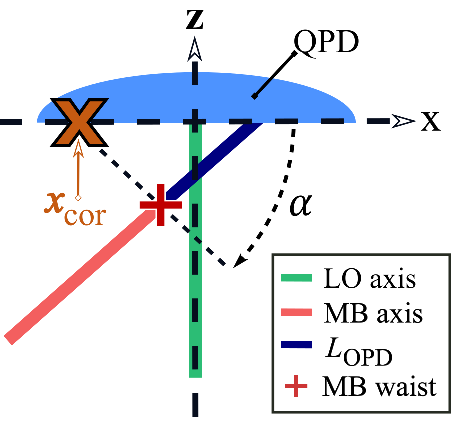}
  \end{center}
  \caption{\label{fig:sketch_lps_x_cor}Test geometry where the measurement beam center of rotation is laterally offset along the $x$-axis to $x_\textrm{cor}$, which is represented as an orange ``$\times$''. The longitudinal center of rotation is still ${z=0}$. These rotations yield additional pathlength $L_\textrm{OPD}$ extending from the new \gls{MB} waist location (the red ``+''). Here, the rotation and offset are negative values.}
\end{figure}

In this test, we laterally shifted the \gls{MB}'s center of rotation ($x_\textrm{cor}$), as illustrated in \cref{fig:sketch_lps_x_cor}. The optical pathlength difference between the \gls{MB} in the tilted and non-tilted case is

\begin{equation}
    \lopdeq(\alpha,\xcor) = \xcor \tan(\alpha) \approx \xcor \alpha
    \,.
    \label{eq:LOPD-Case3}
\end{equation}

\begin{figure}[!htbp]
    \centering \includegraphics[width=.6\textwidth,
    trim=.05cm .05cm .05cm 0cm, clip]{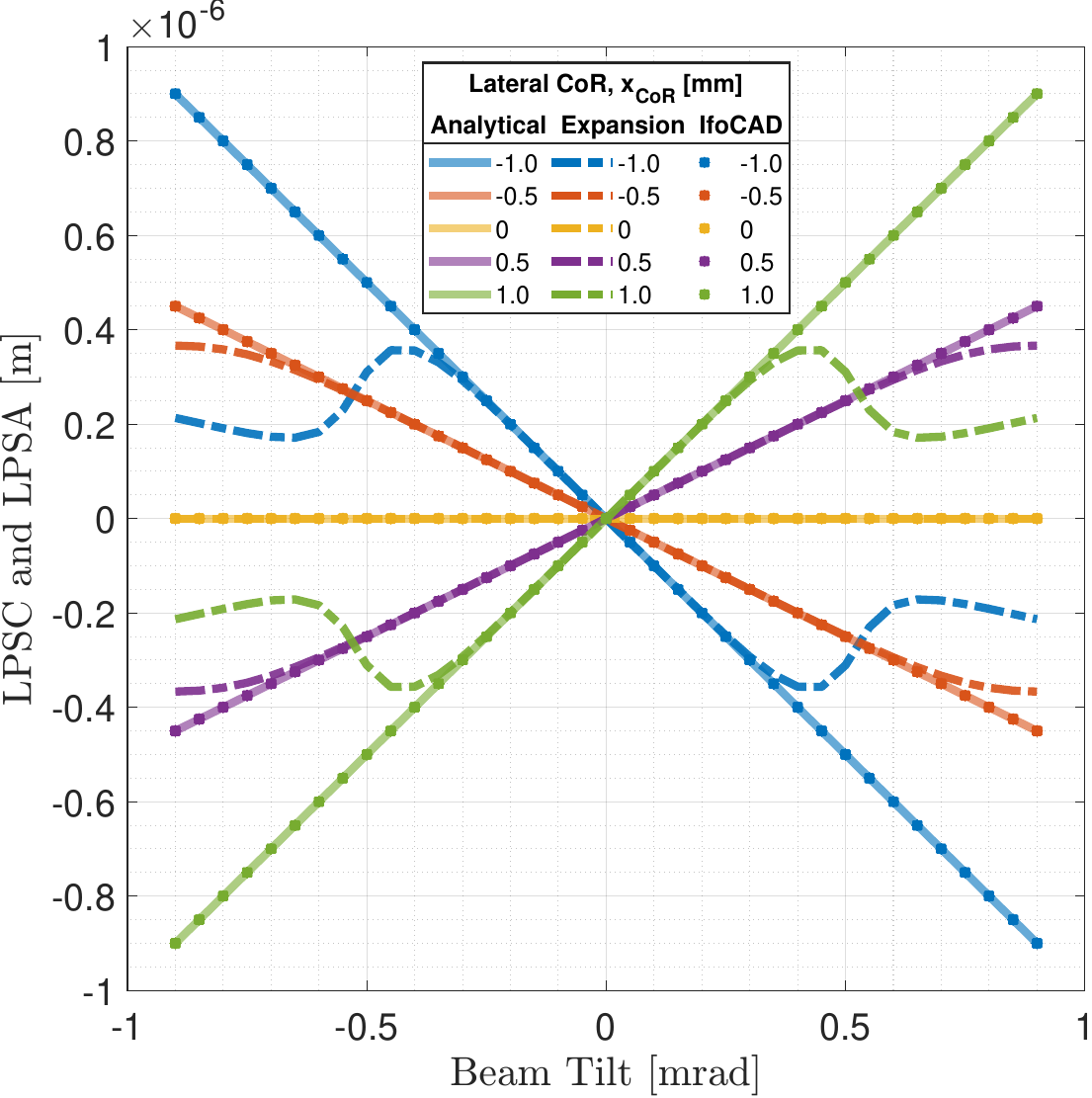} 
    \caption{The plots equivalently show \gls{LPSC} and \gls{LPSA} for a variation to the measurement beam lateral center of rotation ($x_\textrm{cor}$) for values of $\pm \{ 0,\, 0.5, \, 1\}\,$mm.
    Only LPSC values were plotted because differences between the two datasets for all methods were due to numerical error (on the order of $10^{-10}\,$m). The Expansion method shows divergence from the other two methods for large tilt, as expected, due to limitations of the approximation for large static misalignments.}
    \label{fig:lps_x_cor}
\end{figure}

 Notably, this case yields identical results for \gls{LPSA} and \gls{LPSC}, as shown in \cref{fig:lps_x_cor}.
%
The similarity in the two signals is because the \gls{TTL} coupling is dominated by the geometric contribution of the relatively large $\xcor$ variation itself.
Geometrically equivalent to a lateral offset of the \gls{MB}, this generates the same type of lever arm as in $\xoff$ case, thus producing an equivalent (up to sign) geometric \gls{TTL}. 
The functional fitting here was performed again using the Analytical method for the square \glspl{QPD} and \textit{IfoCAD} for the circular, and takes the form:
\begin{equation}
    \Delta \textrm{LPS}(\alpha,\xcor) \approx 
    \lopdeq + \xcor (c_{1,1}\alpha + c_{3,1}\alpha^3)
    \,.
    \label{eq:TTL-Case3}
\end{equation}
The fit coefficients we obtained are listed in \cref{table4}.

\begin{table}[ht!]
 \caption{Fit coefficients for lateral offset in the \gls{MB} center of rotation ($x_\textrm{cor,MB}$).}\label{table4}
 \scaletable{
\begin{tabular}[h!]{c c c c }
\br
    Coefficient &  Small Square QPD Range & Circular QPD Range & Large Square QPD Range\\
    \mr
   LPSC $c_{1,1}$ $\left(\frac{\mathrm{pm}}{(\mathrm{mm})(\muup\mathrm{rad})}\right)$  & $(2.317 \pm 0.021) \times 10^{-4}$ & $(1.991 \pm 0.012) \times 10^{-4}$ & $(1.880 \pm 0.011) \times 10^{-4}$
   \\
LPSA $c_{1,1}$ $\left(\frac{\mathrm{pm}}{(\mathrm{mm})(\muup\mathrm{rad})}\right)$  & $(2.299 \pm 0.003) \times 10^{-4}$ & $(1.977 \pm 0.001) \times 10^{-4}$ & $(1.869 \pm 0.003) \times 10^{-4}$ 
\\
\mr
LPSC $c_{3,1}$ $\left(\frac{\mathrm{pm}}{(\mathrm{mm})(\muup\mathrm{rad}^3)}\right)$  & $(3.492 \pm 0.098) \times 10^{-10}$ & $(3.374 \pm 0.065) \times 10^{-10}$ & $(3.430 \pm 0.050) \times 10^{-10}$
   \\
LPSA $c_{3,1}$ $\left(\frac{\mathrm{pm}}{(\mathrm{mm})(\muup\mathrm{rad}^3)}\right)$  & $(4.560 \pm 0.006) \times 10^{-10}$ & $(4.481 \pm 0.005) \times 10^{-10}$ & $(4.488 \pm 0.009) \times 10^{-10}$
\\
\br
\end{tabular}
}
\end{table}

The relatively large offsets used in this scenario are of particular detriment to the Expansion method - the offset to the \gls{CoR} of $1\,$mm, for instance, is roughly twice the beam size and induces a pathlength difference $\lopdeq\simeq 0.3\,\muup$m at $300\,\muup$rad. 
Regardless, it is clear that the geometric TTL dominates in this case.
This is prominently visible when comparing the linear geometric contribution, defined in \cref{eq:LOPD-Case3}, with the linear non-geometric contribution, $\xcor (c_{1,1}\alpha) $ from \cref{eq:TTL-Case3}. Here, the non-geometric coefficient $c_{1,1}$, with values of approximately $2 \times 10^{-4}$\,pm/mm/$\upmu$rad = $2 \times 10^{-7}$\,$\mathrm{rad}^{-1}$ as listed in \cref{table4}, needs to be compared with a factor $c_{1,1}^\text{geom}$ (which we did not explicitly define in \cref{eq:LOPD-Case3}) of value $c_{1,1}^\text{geom}=$1\,rad$^{-1}$. Thereby, the geometric TTL is dominating in this case by nearly 7 orders of magnitude.

\subsection{Results for longitudinal offset of beam waists}\label{sub:z_off}

\begin{figure}[!htbp]
\begin{center} 
\includegraphics[width=.35\textwidth,
    trim=0 .1cm 0 0, clip]{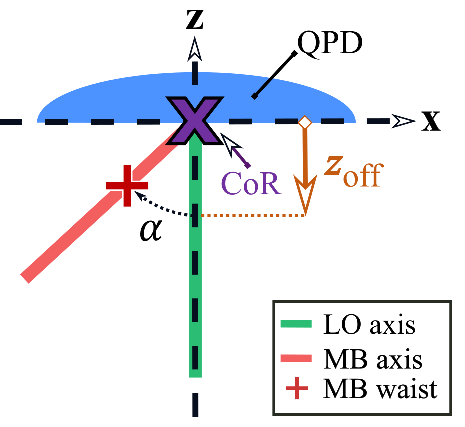}
  \end{center}
  \caption{\label{fig:sketch_lps_z_off}Test geometry where the measurement beam's center of rotation remains the \gls{QPD} center (purple ``$\times$''), but the waist (red ``+'') is longitudinally offset by a distance $z_\textrm{off}$ before rotation in $\alpha$. Note that $L_{OPD}$ is independent of angle and thus zero for this coupling. In this depiction, the rotation and offset are negative values.}
\end{figure}

In this test, we longitudinally shifted the position of the \gls{MB} waist by $z_\textrm{off}$ along its beam axis, as illustrated in \cref{fig:sketch_lps_z_off}.
Here, the additional pathlength from misalignment is a tilt-independent offset which produces a constant phase shift, hence
\begin{equation}
    \lopdeq(\alpha,\zoff) = 0 \,.
\end{equation}
%

\begin{figure}[hb!]
\begin{center}
    \includegraphics[width=.95\textwidth,
    trim=.05cm .05cm .05cm 0cm, clip]{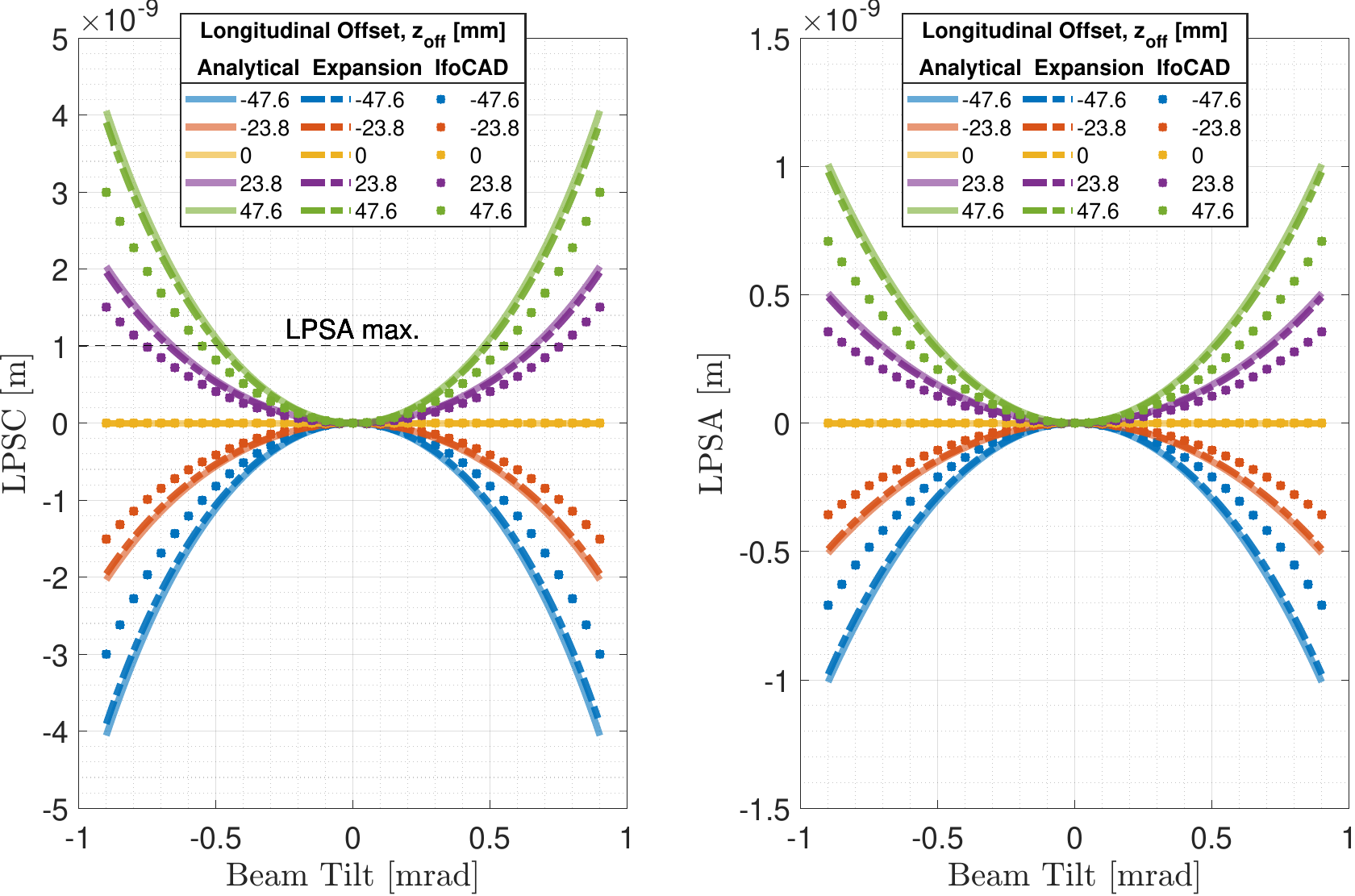}
  \end{center}
  \caption{\label{fig:lps_z_off} The plots show \gls{LPSC} (left) and \gls{LPSA} (right) for large $1\,$mm square \gls{QPD} (``Analytical'', ``Expansion'') or a circular \gls{QPD} (``IfoCAD'') following longitudinal offset of the \gls{MB} waist and tilt of its beam axis about the \gls{QPD} center. In this case, \gls{LPSA} again shows significantly smaller noise coupling than \gls{LPSC}. This is made more evident by the dashed ``LPSA max'' line which denotes the maximal value from the \gls{LPSA} plots.}
\end{figure}

Results for the \gls{LPSC} and \gls{LPSA} signals are shown in \cref{fig:lps_z_off}. 
Clearly, \gls{LPSA} is subject to roughly four times less \gls{TTL} noise than \gls{LPSC} for the case of longitudinal waist position offsets.
Disparity between \textit{IfoCAD} and the two other methods is again due only to difference in \gls{QPD} geometry, which we verified with square \gls{QPD} implementation in \textit{IfoCAD} (cf. \ref{app:square_qpd}). 
 
Because $\lopdeq$ goes to zero here, the total \gls{TTL} contributions from this misalignment coupling can be written as
\begin{equation}\label{eq:zoff_lps}
    \Delta \textrm{LPS}(\alpha,\zoff) \approx \alpha^2\left(c_{2,1} z_\mathrm{off}+c_{2,3}
    z_\mathrm{off}^3\right)
    .
\end{equation}
We present coupling coefficients for this case in \cref{table6}. 
\begin{table}
\caption{Coefficients for TTL effects from MB longitudinal offsets ($z_\textrm{off,MB}$).}\label{table6}
\scaletable{
\begin{tabular}[htb!]{c c c c }
\br
    Coefficient &  Small Square QPD Range & Circular QPD Range & Large Square QPD Range\\
    \br
   LPSC $c_{2,1}$ $\left(\frac{\mathrm{pm}}{(\mathrm{mm})(\muup\mathrm{rad}^2)}\right)$  & $(-3.575 \pm 0.018) \times 10^{-5}$ & $-6.608 \times 10^{-5}$ & $(-8.503 \pm 0.011) \times 10^{-5}$
   \\
LPSA $c_{2,1}$ $\left(\frac{\mathrm{pm}}{(\mathrm{mm})(\muup\mathrm{rad})^2}\right)$  & $(-6.750 \pm 0.038) \times 10^{-6}$ & $(-1.745 \pm 0.004) \times 10^{-5}$ & $(-2.348 \pm 0.018) \times 10^{-5}$ 
\\
\mr
LPSC $c_{2,3}$ $\left(\frac{\mathrm{pm}}{(\mathrm{mm}^3)(\muup\mathrm{rad}^2)}\right)$  & $(1.198 \pm 0.005) \times 10^{-10}$ & $1.872 \times 10^{-10}$ & $(1.888 \pm 0.005) \times 10^{-10}$
   \\
LPSA $c_{2,3}$ $\left(\frac{\mathrm{pm}}{(\mathrm{mm}^3)(\muup\mathrm{rad}^2)}\right)$  & $(2.995 \pm 0.020) \times 10^{-11}$ & $(6.190 \pm 0.001) \times 10^{-11}$ & $(6.779 \pm 0.025) \times 10^{-11}$
\\
\br
\end{tabular}
}
\end{table}

Conversely, if we instead offset the \gls{LO} beam (while still rotating the \gls{MB}) the resulting coefficients are those in \cref{table7}.

\begin{table}
\caption{Coefficients for TTL effects from LO longitudinal offsets ($z_\textrm{off,LO}$).}\label{table7}
\scaletable{
\begin{tabular}[htb!]{ c c c c }
\br
    Coefficient &  Small Square QPD Range & Circular QPD Range & Large Square QPD Range\\
    \br
   LPSC $c_{2,1}$ $\left(\frac{\mathrm{pm}}{(\mathrm{mm})(\muup\mathrm{rad}^2)}\right)$  & $(3.575 \pm 0.018) \times 10^{-5}$ & $6.608 \times 10^{-5}$ & $(8.503 \pm 0.011) \times 10^{-5}$
   \\
LPSA $c_{2,1}$ $\left(\frac{\mathrm{pm}}{(\mathrm{mm})(\muup\mathrm{rad})^2}\right)$  & $(6.750 \pm 0.038) \times 10^{-6}$ & $(1.745 \pm 0.004) \times 10^{-5}$ & $(2.348 \pm 0.018) \times 10^{-5}$ 
\\
\mr
LPSC $c_{2,3}$ $\left(\frac{\mathrm{pm}}{(\mathrm{mm}^3)(\muup\mathrm{rad}^2)}\right)$  & $(-1.198 \pm 0.005) \times 10^{-10}$ & $-1.872 \times 10^{-10}$ & $(-1.888 \pm 0.005) \times 10^{-10}$
\\
LPSA $c_{2,3}$ $\left(\frac{\mathrm{pm}}{(\mathrm{mm}^3)(\muup\mathrm{rad}^2)}\right)$  & $(-2.995 \pm 0.020) \times 10^{-11}$ & $(-6.190 \pm 0.001) \times 10^{-11}$ & $(-6.779 \pm 0.025) \times 10^{-11}$
\\
\br
\end{tabular}
}
\end{table}
It is obvious the coefficients differ between \cref{table6} and \cref{table7} only in sign.
The plots for the case of shifting the \gls{LO} instead are  identical to those of \cref{fig:lps_z_off} except for a sign inversion, but they are not shown here for the sake of brevity.
This indicates that \gls{TTL} noise contributions here depend only on the initial relative longitudinal offset between beam waists. 
Therefore, a more succinct form of \cref{eq:zoff_lps}
for \gls{LPS} contribution in the case of longitudinal offsets is
\begin{equation}
    \Delta \textrm{LPS}(\alpha,\zoff) \approx \alpha^2
    \left(
    c_{2,1}\Delta z+c_{2,3}\Delta z^3
    \right)
    \,
\end{equation}
where $\Delta z = z_\textrm{off,MB}-z_\textrm{off,LO}$ is the distance between beam waists. 

\subsection{Results for longitudinal offset of measurement beam center of rotation}\label{sub:z_cor}

\begin{figure}[hb!]
\begin{center}
    \includegraphics[width=.4\textwidth,
    trim=0 .1cm 0 0, clip]{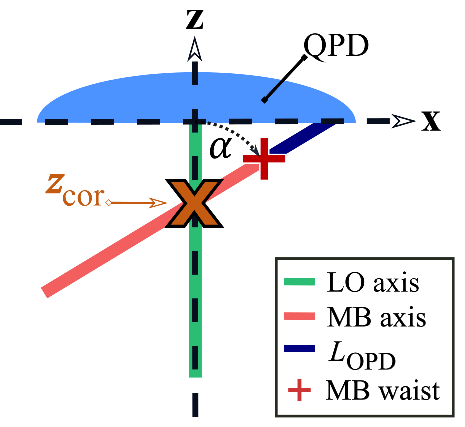}
  \end{center}
  \caption{\label{fig:sketch_lps_z_cor}Test geometry where the measurement beam center of rotation is shifted longitudinally along the $z$-axis by $\zcor$. 
  The orange ``$\times$'' represents the new rotation point.
  These rotations yield additional pathlength $L_\textrm{OPD}$ extending from the new \gls{MB} waist location (the red ``+''). Here, the rotation and offset are both negative values.}
\end{figure}

In this final test, we varied the \gls{MB}'s longitudinal center of rotation ($z_\textrm{cor}$) point along the optical axis, as illustrated by \cref{fig:sketch_lps_z_cor}. The geometric \gls{TTL} contribution here is
\begin{equation}
    \lopdeq = \zcor \left(1-\frac{1}{\cos \alpha}\right)
    \,.
\end{equation}
%

\begin{figure}[htbp!]
    \centering \includegraphics[width=.95\textwidth,
    trim=.05cm .05cm. .05cm 0cm, clip]{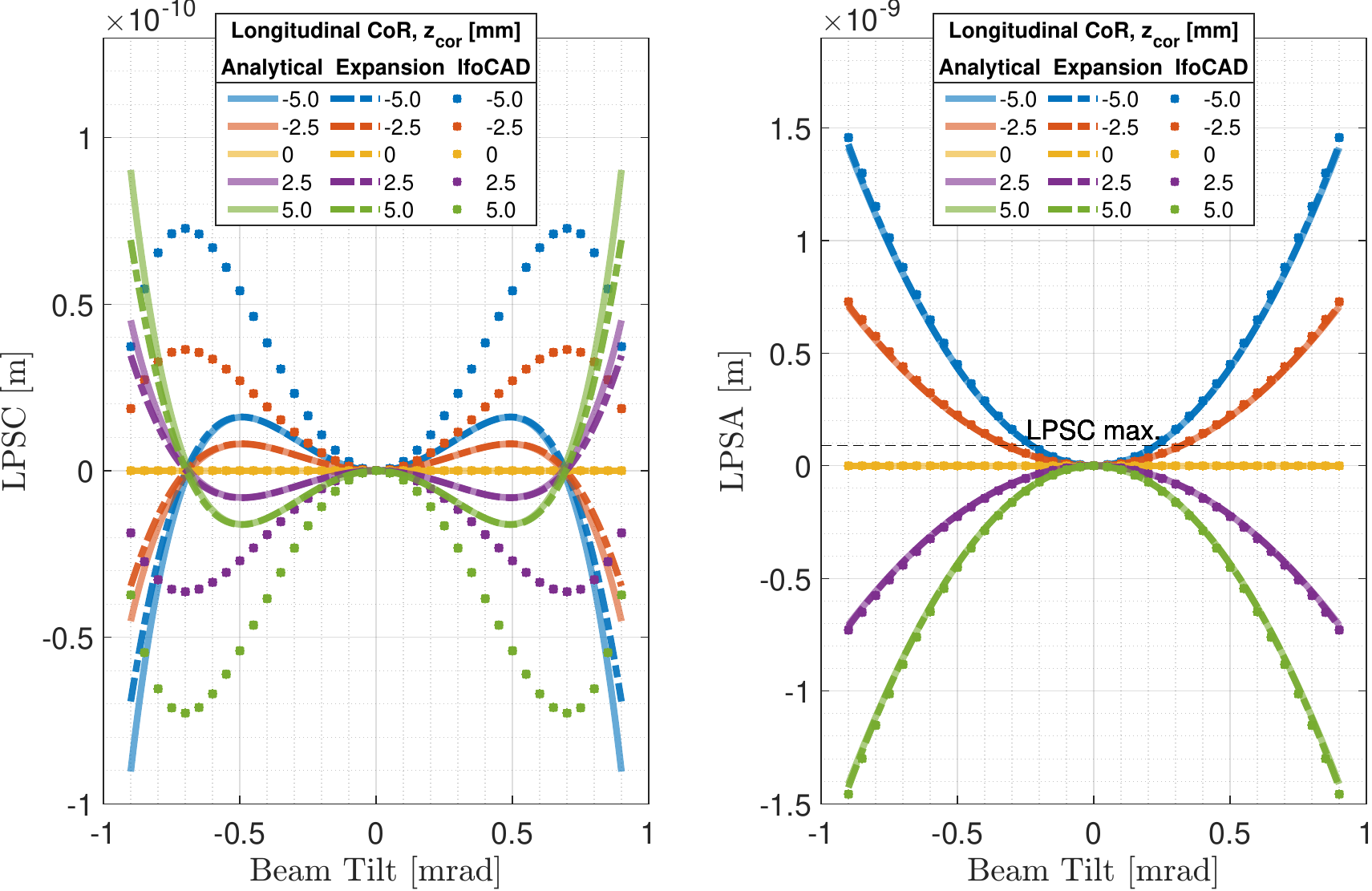} 
    \caption{This figure shows plots of \gls{LPSC} and \gls{LPSA} for variation to the \gls{MB} longitudinal center of rotation ($z_\textrm{cor}$) of $\pm\{0,\,2.5,\,5\}$\,mm for the large square photodiode (``Analytical'',``Expansion'') and circular photodiode (``IfoCAD'').
    This is the single case where \gls{LPSC} shows smaller noise coupling than \gls{LPSA}, which is emphasized by the dashed ``LPSC max'' line denoting the maximal value from the \gls{LPSC} plots.}
    \label{fig:lps_z_cor}
\end{figure}

The \gls{LPS} results for this test are shown in \cref{fig:lps_z_cor}.
Here, we observe a significant difference between the \gls{LPSC} calculated with \textit{IfoCAD}’s circular QPD when compared against the other two methods' large square \gls{QPD} results. 
This disparity is especially apparent beyond $\pm100\,$$\muup $rad in the LPSC.
We verified via square \gls{QPD} implementation in \textit{IfoCAD} that this disparity is solely due to \gls{QPD} geometry (cf. \ref{app:square_qpd}).

The total \gls{TTL} effects can be described accurately via
\begin{equation}
    \Delta \textrm{LPS}(\alpha,\zcor) \approx 
    \lopdeq + z_\mathrm{cor}\left(c_{2,1}\alpha^2 + c_{4,1}\alpha^4\right)
    \;.
\end{equation}
These coefficients are shown in \cref{table5}.
The second-order OPD effect couples to interferometric phase readout and approaches $2\,$nm at $1\,$mrad of beam tilt for $\zcor=5\,$mm.
Ideally, imaging systems in the beam path would mitigate geometric coupling so that the pivot of the beam is re-oriented onto the \gls{QPD} center. 

Displacing the \gls{CoR} along the beam axis is a case of vanishing \gls{TTL} when computed on a large single-element photodiode (SEPD) (cf. \cite{Schuster15_vanishing_ttl}). 
Using the complex-sum formalism to compute the coupling for the signal on a finite-sized \gls{QPD} is almost equivalent to calculating the total phase across the PD surface as a single element, except for some signal power loss at the QPD's slit and edges. 
The \gls{LPSA} formalism does not mimic the signal generation on a large SEPD in the same quality as the \gls{LPSC} formalism does.
Hence, in this specific case, we expect the coupling in the \gls{LPSC}  to be less than in the \gls{LPSA}, and this is what can be seen in \cref{fig:lps_z_cor}.
Therefore, this case is the first and only one of our 5 cases where the \gls{LPSC} is less susceptible to TTL coupling than the \gls{LPSA} formalism.

\begin{table}
\caption{Coefficients for longitudinal offset in MB center of rotation ($z_\textrm{cor,MB}$).}\label{table5}
\scaletable{
\begin{tabular}[htb!]{ c c c c }
\br
    Coefficient &  Small Square QPD Range & Circular QPD Range & Large Square QPD Range\\
    \br
   LPSC $c_{2,1}$ $\left(\frac{\mathrm{pm}}{(\mathrm{mm})(\muup\mathrm{rad})^2}\right)$  & $(-6.484 \pm 0.020) \times 10^{-4}$ & $(-5.576 \pm 0.016) \times 10^{-4}$ & $(-5.278 \pm 0.015) \times 10^{-4}$
   \\
LPSA $c_{2,1}$ $\left(\frac{\mathrm{pm}}{(\mathrm{mm})(\muup\mathrm{rad}^2)}\right)$  & $(-8.997 \pm 0.001) \times 10^{-4}$ & $(-8.603 \pm 0.001) \times 10^{-4}$ & $(-8.487 \pm 0.003) \times 10^{-4}$ 
\\
\mr
LPSC $c_{4,1}$ $\left(\frac{\mathrm{pm}}{(\mathrm{mm})(\muup\mathrm{rad}^4)}\right)$  & $(5.754 \pm 0.746) \times 10^{-11}$ & $(5.492 \pm 0.613) \times 10^{-11}$ & $(5.712 \pm 0.595) \times 10^{-11}$
   \\
LPSA $c_{4,1}$ $\left(\frac{\mathrm{pm}}{(\mathrm{mm})(\muup\mathrm{rad}^4)}\right)$  & $(3.781 \pm 0.052) \times 10^{-12}$ & $(7.490 \pm 6.710) \times 10^{-13}$ & $(4.930 \pm 7.620) \times 10^{-13}$
\\
\br
\end{tabular}
}
\end{table}

\subsection{Removing linear tilt-to-length coupling by applying a model fit}
\label{sub:non-linear-TTL}

In \cref{sub:x_off,sub:x_cor,sub:z_cor,sub:z_off}, we have shown four instances of \gls{TTL} coupling caused by misalignment in the initially defined setup.
Some of our resulting plots of the longitudinal pathlength signals show notably non-linear \gls{TTL} coupling across the examined beam angle range (i.e., $ \pm \SI{900}{\upmu rad}$). 
In practice, the SC's angular jitter will be at the \SI{}{nrad/\sqrt{Hz}} level, accompanied by a µrad-level angular offset (referred to as a DC offset). 
Thereby, the jitter will probe the nonlinear curve in a very small range around some point.
Consequently, the \gls{TTL} coupling expected in \gls{LISA} will resemble a dominantly first order Taylor expansion of the presented curves, and it will be centered around the DC offset angle. 
Hence, one possible mitigation scheme involves fitting the \gls{TTL} coupling with a linear model and removing this component from the data in the post-processing, as discussed in the works of \cite{don_paper,post_subtraction}.

In this subsection, we investigate the non-linear \gls{TTL} residuals of our simulated pathlength signals after this linear subtraction scheme.
In \cite{post_subtraction}, for instance, an approximately white jitter of $5\,$\SI{}{nrad/\sqrt{Hz}} is projected for \gls{LISA}  in roughly the \SIrange{3}{100}{mHz} band.
We use this $5\,$nrad jitter value for testing the relevance of the non-linear TTL coupling of the \gls{TMI}. 
Here, the beam tilt seen at the \gls{QPD} level is scaled by a factor of five (2 for the TM round-trip, and 2.5 for the magnification factor (cf.~\cref{Sec:SimSetup})).
Thus, we expect the \gls{MB} tilt of our test cases to probe around a $\pm 25\,$nrad region, centered at some DC offset.
We assume such offsets on the order of 10s of \SI{}{\upmu rad} to be realistic for the LISA \glspl{TMI}.

Under these considerations, our implementation of a linear \gls{TTL} subtraction follows these steps:
\begin{itemize}
    \item We first apply a fourth order polynomial fit to our results of both \gls{LPSA} and \gls{LPSC} for each misalignment test case.
    \item We then linearize these fits in $\pm 25\,$nrad beam jitter regions, which are each centered at $50 \, \muup$rad steps of our $\pm 900\,$\SI{}{\upmu rad} beam tilt range. 
    This allows us to probe the signals around various DC offset angles in approximately linear regimes.
    \item Finally, we compute length residuals when the linear fit is subtracted off the fourth-order fit function.
\end{itemize}

\begin{figure}
    \centering
    \includegraphics[width=0.95\linewidth]{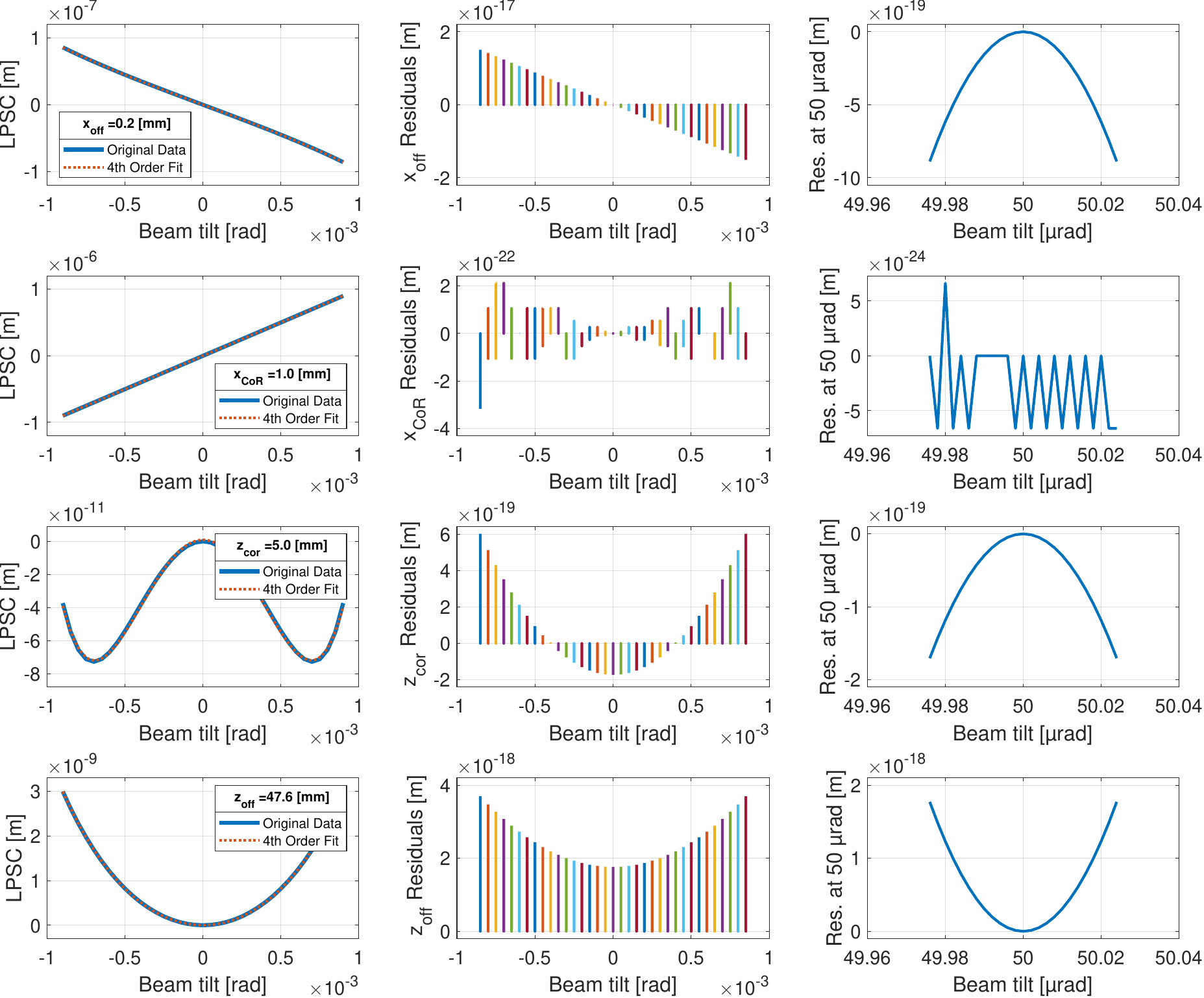}
    \caption{Plots of \gls{LPSC} fits and residuals for each misalignment case (from \textit{IfoCAD} circular QPD data) as explained in the text. 
    }
    \label{fig:enter-label1}
\end{figure}

\begin{figure}
    \centering
    \includegraphics[width=0.95\linewidth]{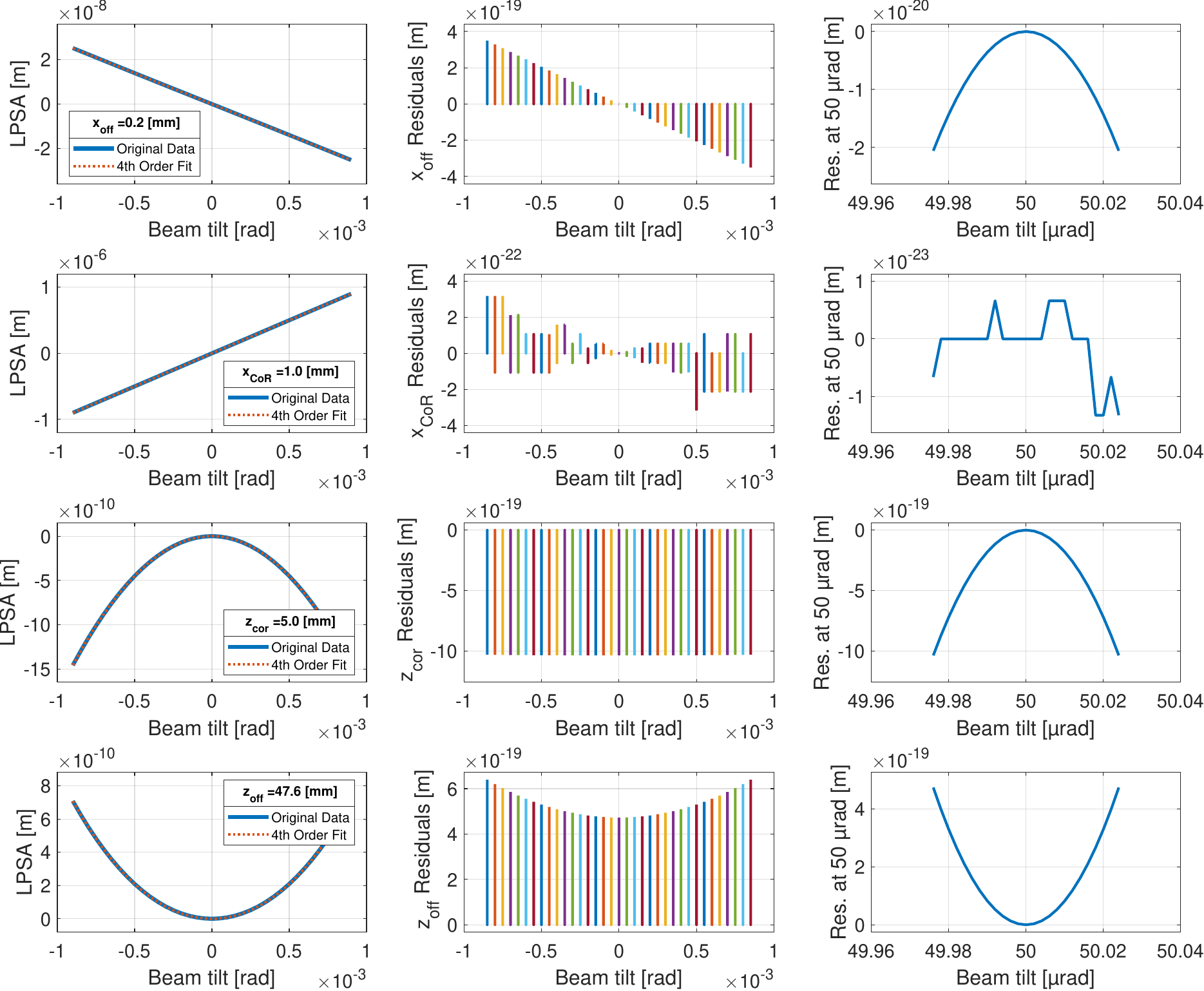}
    \caption{Plots of \gls{LPSA} fits and residuals for each misalignment case (from \textit{IfoCAD} circular QPD data) as explained in the text. 
    }
    \label{fig:enter-label2}
\end{figure}

The plots in \cref{fig:enter-label1} and \cref{fig:enter-label2} show the corresponding residuals in both signals (\gls{LPSC} and \gls{LPSA}, respectively) for the largest test value in each type of misalignment. 
These data are computed from the \textit{IfoCAD} circular \gls{QPD} results.
In the figures, each row shows plots for one particular misalignment.
The first column shows the fourth order fit and its good overlap with the original data. 
The second column shows the residuals at each of the $\pm 25\,$nrad linear fit regions. 
Due to the very small x-range we used for each colored line, the residuals appear as vertical lines, but they are polynomial functions (except for numerical noise).
This is made apparent in the third column, which shows ``zoomed-in'' plots of each fit, at the $\sim$nrad scale. 
%
%
These plots show very optimistically that the residuals following this subtraction scheme are $\mathcal{O}$($10^{-22}$ to $10^{-17}\,$m), far below the picometer-level criteria required for LISA.
Hence, the non-linear contributions to \gls{TTL} noise for all of the cases we have tested in this work are negligible under the assumptions of our subtraction scheme.

%% file: S5_Conclusions.tex
\section{Summary and conclusion}\label{Section5}

We have shown case-by-case simulations of the tilt-to-length coupling for the LISA \gls{TMI} in two types of signal formulations: the averaged phase (LPSA) and the complex sum (LPSC). 
For all cases, we have tested three different methods, and have found good agreement of the results, particularly if identical photodiode geometries were used.

In total, we have tested five different cases of static misalignments.
In the first case, we tested waist size mismatches (\cref{sub:w0_diff}), which led to vanishing \gls{TTL} in both signal types.
We also tested two cases of misalignment in which \gls{LPSA} showed less \gls{TTL} coupling than \gls{LPSC}.
These were the case of lateral offset of the \gls{MB} (or \gls{LO}) beam axis (\cref{sub:x_off}) and the case of longitudinal offset of the \gls{MB} waist location (\cref{sub:z_off}).
A fourth case, lateral offset of the \gls{MB} center of rotation (\cref{sub:x_cor}), proved to be dominated by geometric \gls{TTL} coupling so that the two formulations performed equally.  

In the fifth case, we investigated variation to the \gls{MB} longitudinal center of rotation (\cref{sub:z_cor}).
This is an exceptional case, where a large SEPD would not sense any TTL coupling.
The LPSC method is the best approximation of the longitudinal pathlength signal of an SEPD obtainable from a QPD. 
Therefore, the LPSC shows the best approximation of the SEPD's zero-coupling, and, therefore, less TTL coupling than the LPSA signal.

For LISA it is currently expected that the \gls{TTL} coupling will be reduced by subtracting a linear fit model from the measured LPS data. 
The magnitude of the TTL coupling coefficients prior to subtraction influences the residual noise level obtained after this subtraction is performed. 
Therefore, the choice of the \gls{LPS} formalism might indeed have an impact on the final noise performance.
From the results we have presented here, it appears that the \gls{LPSA} formulation is the favorable choice for the \gls{TMI}.
However, the \gls{TMI} is known to be subject to significantly less \gls{TTL} noise than the \gls{ISI}. 
Therefore, an extension of this study to the \gls{ISI} is needed to finally judge which signal formalism is a better choice there for TTL suppression. 

Finally, we investigated the non-linear \gls{TTL} contributions in the \gls{TMI}.
For this, we subtracted a linear fit model from our LPS-data and tested the magnitude of the coupling for jitter levels currently expected for LISA. 
We found the non-linear TTL contributions to be negligible in all of our test cases of misalignments in the TMI relative to LISA requirements.

%% file: A1_square_qpd_check.tex
\section{Verification of the IfoCAD square photodiode geometry}\label{app:square_qpd}

Plots of longitudinal pathlength signals of \cref{Section4} only showed the IfoCAD method results for the circular \gls{QPD} geometry.
However, disparities existed between these results and the two other methods which were particularly significant for certain misalignments.
These were the $\xoff$ (\cref{sub:x_off}), $\zoff$ (\cref{sub:z_off}), and $\zcor$ (\cref{sub:z_cor}) cases, which we now present here as verification that all three methods very closely agree when the same geometry is applied.
All of these plots show \gls{LPSC} and \gls{LPSA} results for only the $1\,$mm width square \gls{QPD}.
%
\begin{figure}[!htbp]
    \centering \includegraphics[width=.95\textwidth,
    trim=4cm 1cm 2cm 1cm, clip]{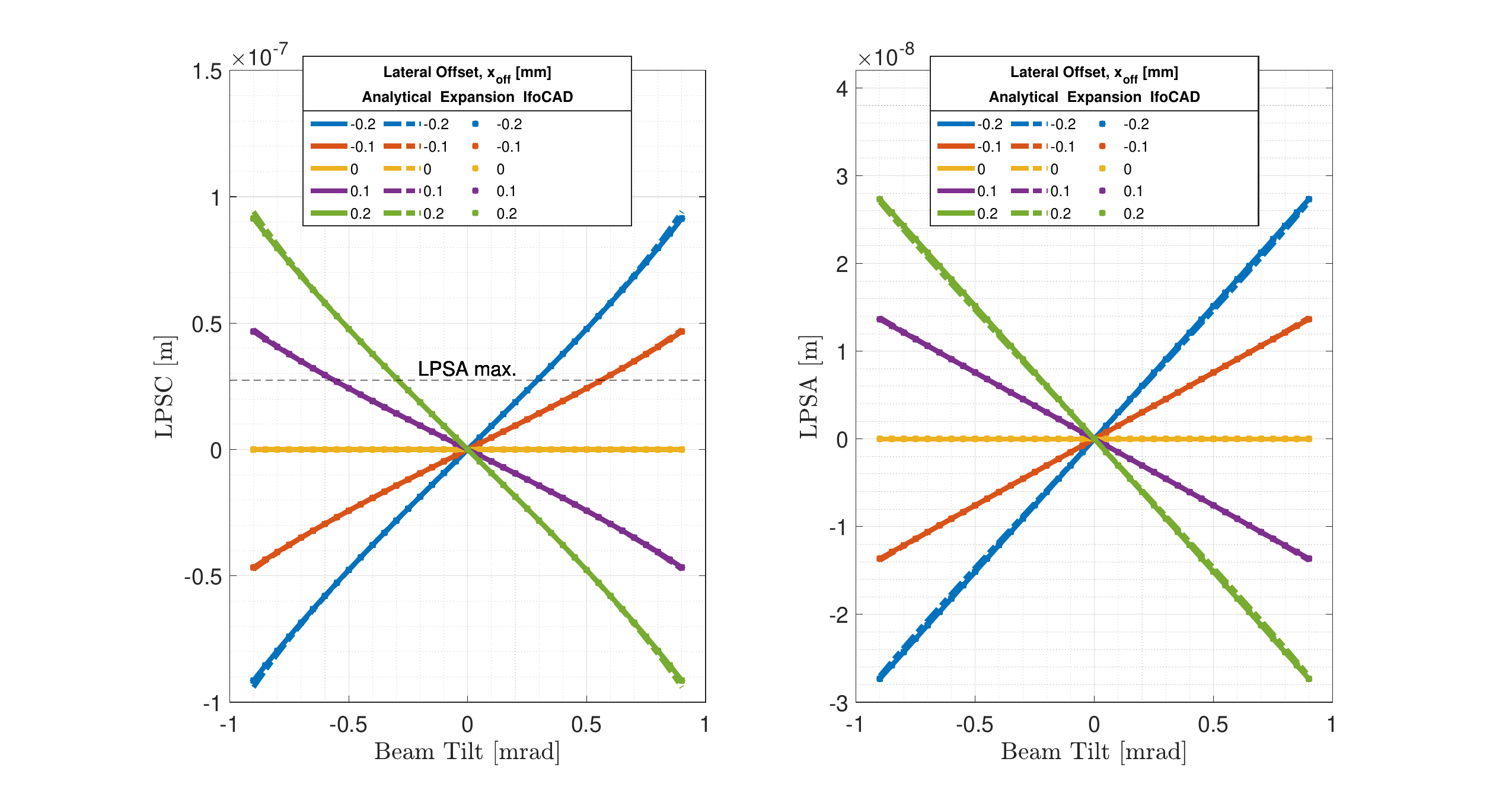} 
    \caption{Variation to the measurement beam lateral center of rotation ($x_\textrm{off}$), where all three methods use a $1\,$mm width square \gls{QPD} (cf. \cref{fig:1mm_x_off}).
    }\label{fig:appendix_xoff}
\end{figure}
%
\begin{figure}[!htbp]
    \centering \includegraphics[width=.95\textwidth,
    trim=.05cm .05cm .05cm 0cm, clip]{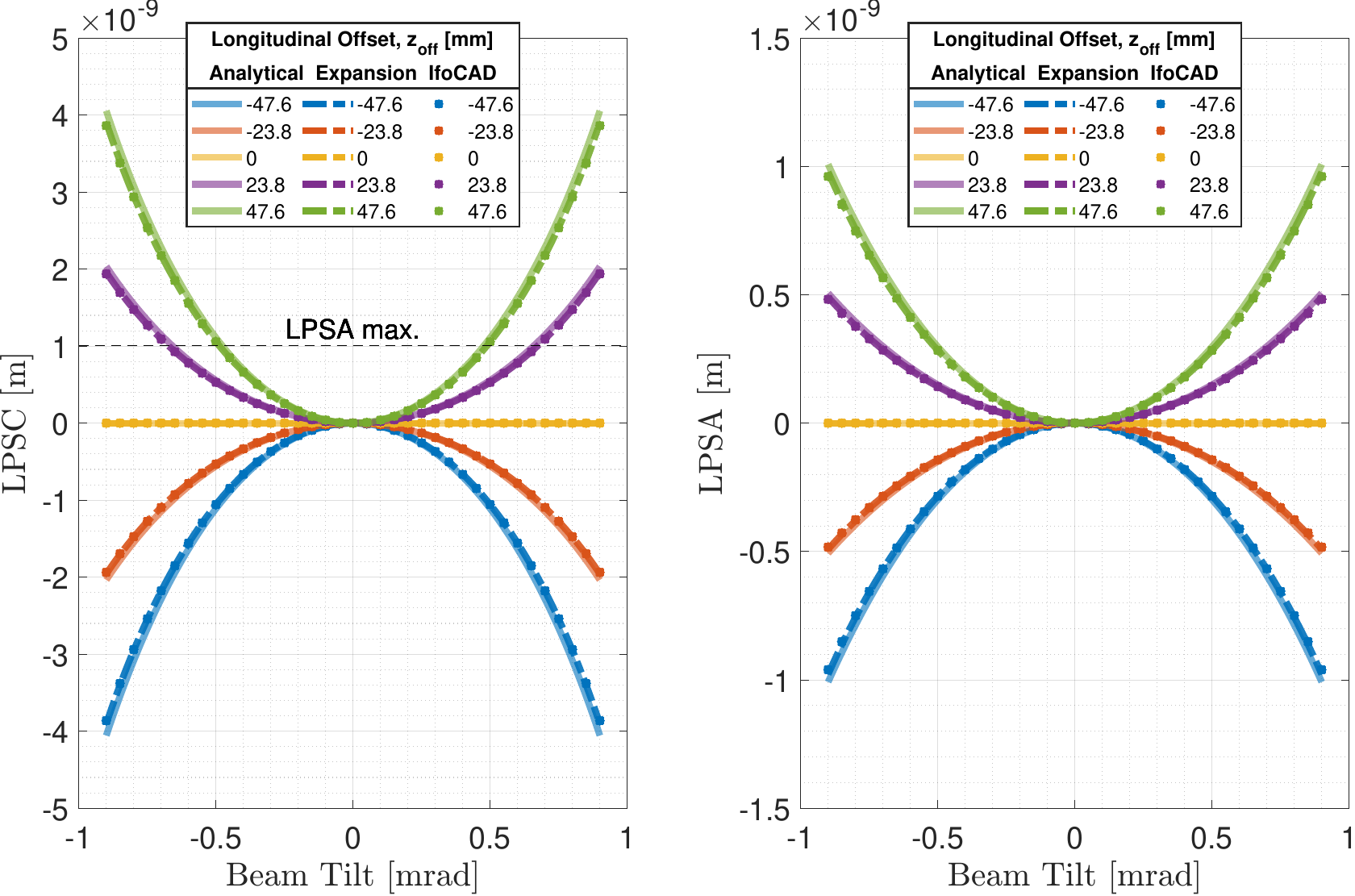} 
    \caption{Variation to the measurement beam waist offset ($\zoff$), where all three methods use a $1\,$mm width square \gls{QPD} (cf. \cref{fig:lps_z_off}).} \label{fig:appendix_zoff}
\end{figure}
%
\begin{figure}[!htbp]
    \centering \includegraphics[width=.95\textwidth,
    trim=.05cm .05cm .05cm 0cm, clip]{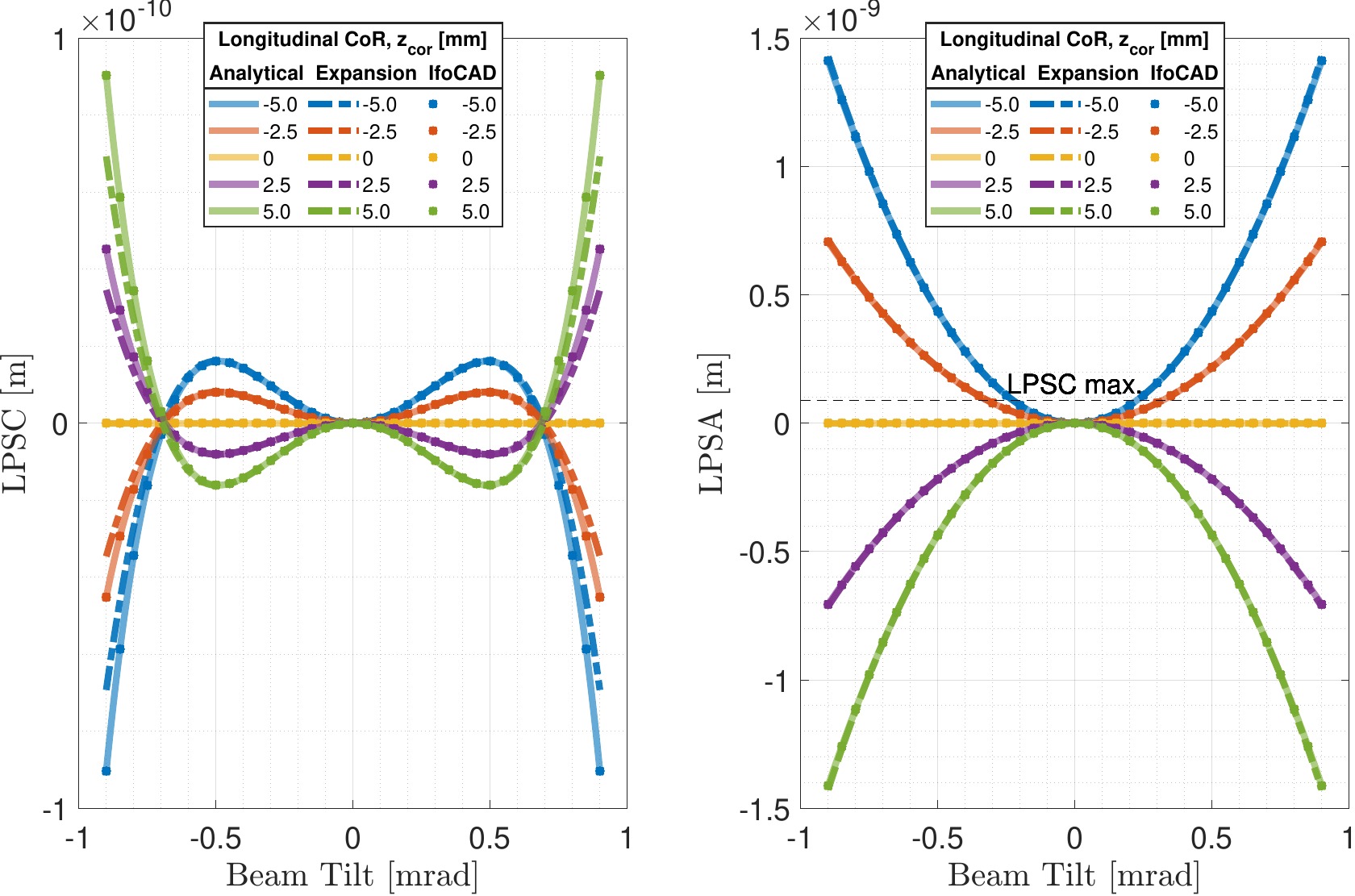} 
    \caption{Variation to the measurement beam longitudinal center of rotation ($\zcor$), where all three methods use a $1\,$mm width square \gls{QPD} (cf. \cref{fig:lps_z_cor}).}\label{fig:appendix_zcor}
\end{figure}
\clearpage